\documentclass[aps,10pt,twocolumn,showpacs,citeautoscript,amsmath,amssymb,floatfix,superscriptaddress]{revtex4-1}

\usepackage{graphicx}
\usepackage{dcolumn}
\usepackage{bm}
\usepackage{epsfig,hyperref,amsthm}
\usepackage{mathtools}
\usepackage{color}
\usepackage{colordvi}
\usepackage[dvipsnames]{xcolor}
\usepackage{relsize}
\usepackage{accents}
\usepackage{wasysym}  
\usepackage[caption=false]{subfig}

\hypersetup{
	colorlinks=true,
	linkcolor=blue,
	citecolor=blue,
	urlcolor=blue
}

\newcommand{\ave}[1]{\left< #1 \right>}
\newcommand{\ket}[1]{\ensuremath{|#1\rangle}}
\newcommand{\bra}[1]{\ensuremath{\langle #1|}}
\newcommand{\braket}[2]{\langle #1|#2\rangle}
\newcommand{\ketbra}[2]{|#1 \rangle  \langle #2 |}
\newcommand{\proj}[1]{\ket{#1}\bra{#1}}
\newcommand{\be}{\begin{equation}}
\newcommand{\ee}{\end{equation}}
\newcommand{\ba}{\begin{eqnarray}}
\newcommand{\ea}{\end{eqnarray}}
\newcommand{\n}{\nonumber}
\newcommand{\lan}{\langle}
\newcommand{\ran}{\rangle}
\newcommand{\rl}{\rangle \langle}
\newcommand{\B}{\beta}
\newcommand{\A}{\alpha}
\newcommand{\ho}{\hat{\mathcal{O}}}
\newcommand{\pe}{\Lambda}
\newcommand{\Des}[1]{\mathcal{B}_{#1}}
\newcommand{\Desd}{\Des{\mathbb{C}^d}}
\newcommand{\Desf}{\Des{\mathbb{C}^d}^{\rm bound}}
\newcommand{\tket}[1]{\ket{\overline{#1}}}
\newcommand{\tbra}[1]{\bra{\overline{#1}}}
\newcommand{\tE}[1]{\overline{E}_{#1}}
\newcommand{\tproj}[1]{\ket{\overline{#1}}\bra{\overline{#1}}}
\newcommand{\tp}[1]{{P}_{#1}}
\newcommand{\tpx}[1]{\overline{P}^{\rm X}_{#1}}
\newcommand{\tqx}[1]{\overline{Q}^{\rm X}_{#1}}
\newcommand{\te}{\tilde{\mathcal{E}}}
\newcommand{\fp}{\mathcal{F}_+}
\newcommand{\da}{\delta_{\rm A}}
\newcommand{\db}{\delta_{\rm B}}
\newcommand{\ka}{\kappa_{\rm A}}
\newcommand{\kb}{\kappa_{\rm B}}
\newcommand{\defeq}{\mathrel{\mathop:}=}
\newcommand{\eqdef}{=\mathrel{\mathop:}}
\newcommand{\E}{\mathcal{E}}

\newcommand{\beq}{\begin{equation}}
\newcommand{\eeq}{\end{equation}}
\newcommand{\beqar}{\begin{eqnarray}}
\newcommand{\eeqar}{\end{eqnarray}}

\newcommand{\tr}[2]{\mathrm{Tr}_{#1}\left[ #2 \right]}
\newcommand{\bip}{\mathbb{C}^d\otimes\mathbb{C}^d}
\newcommand{\bipstate}{\mathcal{L}(\mathbb{C}^d\otimes\mathbb{C}^d)}
\newcommand{\norm}[1]{\left\|#1\right\|}
\newcommand{\sbr}[1]{\left[ #1 \right]}
\newcommand{\lbr}[1]{\left\{ #1 \right\}}
\newcommand{\rIso}{\rho_{\rm iso}}
\newcommand{\id}{\mathbb{I}}
\newcommand{\sys}{{\bf S}}
\newcommand{\env}{{\bf E}}
\newcommand{\bt}{{\bf B}_T}
\newcommand{\obs}{{\bf O}}
\newcommand{\pro}{\mathcal{P}}
\newcommand{\act}{\mathcal{A}}
\newcommand{\FEF}{{\rm FEF}}
\newcommand{\F}[1]{\mathcal{F}_{\rm max}(#1)}
\newcommand{\Sab}[1]{S(A|B)_{#1}}
\newcommand{\Sba}[1]{S(B|A)_{#1}}
\newcommand{\We}[1]{W_{\rm Er}(#1)}
\newcommand{\Wt}[1]{W_{\rm Total}(#1)}
\newtheorem{theorem}{Inequality}

\newtheorem{question}{Question}

\newcommand{\Q}[1]{{\begin{center}\color{purple}\fbox{\begin{minipage}{.95\columnwidth}\small {\color{purple}\begin{question}{\rm #1}\end{question}}\end{minipage}}\end{center}}}
\definecolor{nred}{rgb}{0.9,0.1,0.1}
\definecolor{nblack}{rgb}{0,0,0}
\definecolor{nblue}{rgb}{0.2,0.2,0.8}
\definecolor{ngreen}{rgb}{0.2,0.6,0.2}
\definecolor{ublue}{rgb}{0,0,0.5}
\definecolor{pur}{rgb}{0.75,0,0.75}
\definecolor{nngrn}{rgb}{0,0.5,0.5}
\newcommand{\red}{\color{nred}}
\newcommand{\blu}{\color{nblue}}
\newcommand{\grn}{\color{ngreen}}
\newcommand{\blk}{\color{nblack}}
\newcommand{\sBox}[1]{
{\begin{center}\fbox{\begin{minipage}{.95\columnwidth}\small \blu #1\end{minipage}}\end{center}}}
\newcommand{\Co}{{\rm C}}
\newcommand{\Ho}{{\rm H}}
\newcommand{\ptpm}{p^{\rm TPM}}
\newcommand{\pwm}{p^{\rm W}}
\newcommand{\ptwm}{\tilde{p}^{\rm W}}
\newcommand{\qwm}{Q^{\rm W}}
\newcommand{\pttpm}{\tilde{p}^{\rm TPM}}
\newcommand{\qt}{Q(\tilde{U})}
\newcommand{\qttpm}{Q^{\rm TPM}(\tilde{U})}

\makeatother

\newcommand{\veryshortarrow}[1][3pt]{\mathrel{%
   \hbox{\rule[\dimexpr\fontdimen22\textfont2-.2pt\relax]{#1}{.4pt}}%
   \mkern-4mu\hbox{\usefont{U}{lasy}{m}{n}\symbol{41}}}}
   
\newcommand{\U}{\mathcal{U}}
\newcommand{\Ud}{\mathcal{U}^{\dagger}}
\newcommand{\qtpm}{Q^{\rm TPM}}
\newcommand{\pii}{\Pi^{i_\Co i_\Ho}}
\newcommand{\pif}{\Pi^{f_\Co f_\Ho}}
\newcommand{\itf}{_{i_\Co i_\Ho \veryshortarrow  f_\Co f_\Ho}}
\newcommand{\fti}{_{f_\Co f_\Ho \veryshortarrow i_\Co i_\Ho }}
\newcommand{\Tr}{{\rm Tr}}

\newcommand{\ud}{U^{\dagger}}
\newcommand{\ut}{\tilde{U}}
\newcommand{\utd}{\tilde{U}^{\dagger}}

\newcommand{\otz}{_{10,01}}
\newcommand{\zto}{_{01,10}}

\begin{document}
\title{A quasiprobability distribution for heat fluctuations in the quantum regime}

\author{Amikam Levy}

\affiliation{Department of Chemistry, University of California, Berkeley, Berkeley, California 94720,
	United States}

\affiliation{The Raymond and Beverly Sackler Center for Computational Molecular and Materials Science, Tel Aviv University, Tel Aviv, Israel 69978}

\affiliation{Department of Chemistry, Bar-Ilan University, Ramat-Gan 52900, Israel}

\author{Matteo Lostaglio}
\thanks{The authors contributed equally to this work}

\email{lostaglio@gmail.com, \quad amikamlevy@gmail.com.}
\affiliation{ICFO-Institut de Ciencies Fotoniques, The Barcelona Institute of Science and Technology, Castelldefels (Barcelona), 08860, Spain}
\affiliation{QuTech, Delft University of Technology, P.O. Box 5046, 2600 GA Delft, The Netherlands}

\date{\today}

\begin{abstract}
The standard approach to deriving fluctuation theorems fails to capture the effect of quantum correlation and coherence in the initial state of the system. Here we overcome this difficulty  and derive heat exchange fluctuation theorem in the full quantum regime by showing that the  energy exchange between two locally thermal states in the presence of initial quantum correlations is faithfully captured by a quasiprobability distribution. Its negativities, being associated with proofs of contextuality, are proxys of non-classicality. We discuss the thermodynamic interpretation of  negative probabilities, and provide  \emph{heat flow inequalities} that can only be violated in their presence.
 Remarkably, testing these fully quantum inequalities, at arbitrary dimension, is not more difficult than testing traditional fluctuation theorems.
 We test these results on data collected in a  recent experiment studying the heat transfer between two qubits, and give examples for the capability of witnessing negative probabilities at higher dimensions.

\end{abstract}

\maketitle

\section{Introduction}
Consider two systems, C for `cold' and H `hot', in thermal states at temperatures $T_\Co < T_\Ho$. If the overall system is isolated, energy is conserved, and there are no initial correlations between C and H, heat will flow on average  from the hot to the cold body: 
\begin{equation}
\label{eq:uncorrelatedbound}
Q \equiv Q(\Ho \rightarrow \Co) \leq 0,
\end{equation}
as mandated by the second law of thermodynamics. Eq.~\eqref{eq:uncorrelatedbound} can be derived for both classical and quantum systems  and understood as an average over heat fluctuations that satisfy Jarzynski's exchange fluctuation theorem (XFT)~\cite{jarzynski2004classical}.

The situation is less straightforward if the initial state is locally thermal but correlated, since correlations allow for temporary  `backflows', i.e. $Q > 0$. \cite{lloyd1989use, jennings2010entanglement}.  However, as long as the system is simply \emph{classically} correlated, we can still understand these flows as emerging from averages of underlying classical fluctuations~\cite{jennings2012exchange}. 

This picture breaks down in the presence of quantum correlations. In quantum theory, access to the energy of the system, necessary for defining fluctuations classically, can be acquired only via a measurement process that destroys the very quantum properties we wish to investigate. A striking example is given by \emph{strong heat backflows}, i.e. the observation of an amount of heat flowing from C to H larger than $(\Delta \beta)^{-1} \log d$, where \mbox{$\Delta \beta = \beta_\Co - \beta_\Ho$, $\beta_{\rm X} = 1/(kT_{\rm X})$}, with $k$ Boltzmann's constant and $d$ the local dimension. Observation of such flows implies that the underlying state is entangled~\cite{jennings2010entanglement}. 
If we attempt to understand this flow as originating from  underlying (classical) heat fluctuations, the energy measurements required to access them destroy the entanglement and so the effect disappears. In fact, any stochastic interpretation of heat flows that reproduces the observed $Q$ faces severe no-go theorems~\cite{perarnau2017no, lostaglio2018quantum}. This is the reason why the leading proposal for defining and measuring heat and work fluctuations, the `two-projective-measurement' (TPM), clearly cannot reproduce quantum effects such as \textit{strong} heat  backflows.

 We propose that \emph{quantum} heat fluctuations should be associated with a \emph{quasiprobability} distribution. This arises naturally from a correction to the TPM distribution and can be accessed by weak measurements. We show that the associated negative `probabilities' have a clear thermodynamic interpretation as contributions to the heat flows which cannot be explained within the framework of stochastic thermodynamics. These quasiprobabilities satisfy an XFT that incorporates quantum correlations in the initial state and includes in the relevant classical limits the original XFT of Ref.~\cite{jarzynski2004classical} as well as its extension to classically correlated systems of Ref.~\cite{jennings2012exchange}.  More importantly, our results provide heat flow inequalities that can only be violated in the presence of negativities. These violations are a proxy for a strong form of nonclassicality known as contextuality and remarkably can be evaluated by means of TPM alone.

The challenges posed by the construction of a theory of heat fluctuations reproducing the observed  flows are a strong indication that certain flow configurations are inherently quantum mechanical. We propose that quasiprobabilities, a tool already successfully applied in fields such as quantum optics and quantum computing, can be important diagnostic tools for exploring quantum effects in thermodynamics as our ability to measure heat flows improves.

In Sec.~\ref{sec:problem} we present the problem of capturing quantum phenomena in fluctuation theorems and introduce the concept of quasiprobabilities as the solution. 
In Sec.~\ref{sec:quasi}
we introduce a quasiprobability distribution for heat fluctuation and discuss its thermodynamic interpretation, its quantum nature, and the classical limit. 
Sec.~\ref{sec:qubits} focuses on the two qubit system, and presents heat flow inequalities that witness negativity. We further demonstrate the result on recent experimental data.
In Sec.~\ref{sec:qxft} we present the full quantum heat exchange fluctuation theorem (QXFT), and extend the results on heat flow inequalities to arbitrary dimension systems. In Sec.~\ref{sec:highD} we show that witnessing negativity in higher dimensional systems is possible using  projective energy measurement alone, and we demonstrate this on a two qutrit system. Finally in Sec.~\ref{sec:outlook} we summarize and discuss future possible inquires.   

\subsection*{Comparison with previous works}

 In recent years several proposals have attempted to go beyond some of the limitations faced by TPM schemes. In \cite{deffner2016quantum} a modified quantum Jarzynski equality is derived, which accounts for the thermodynamic cost of the measurement. This in turn led to approaches based on an (initial) one-time measurement scheme for closed \cite{beyer2020work} and open \cite{sone2020quantum} systems, which solve experimental and conceptual difficulties related to performing the second projective measurement in the TPM scheme. Here, on the other hand, we want to	confront the loss of information about quantum effects caused by the first projective measurement of the TPM scheme.
	
	The definition of fluctuations in the quantum regime is not unique. There is in fact a zoo of such proposals that reduce to the TPM scheme in the case of initial states diagonal in the energy basis, and whose first moments coincide with the average energy change (of the undisturbed system), see review~\cite{baumer2018}. 
These proposals necessarily fall into one of two categories~\cite{perarnau2017no, lostaglio2018quantum}:
	\begin{enumerate}
		\item \emph{Lack of positivity:} approaches that give up the idea of assigning probabilities to fluctuations, by allowing `negative probabilities' (Margenau-Hill quasiprobability~ \cite{allahverdyan2014nonequilibrium}, full counting statistics \cite{solinas2015full, solinas2016probing, xu2018effects}, consistent histories \cite{miller2016time, goldstein1995linearly}). These approaches call for an interpretation of such negativities.
		\item \emph{Lack of $\sigma$-additivity:} approaches based on probabilities, which give up the linearity in the density operator of the initial state (Hamilton-Jacobi framework \cite{sampaio2017impossible}, quantum 
		Bayesian network \cite{micadei2020quantum}, one-time final energy measurement \cite{gherardini2020role}). For example, if we consider two preparations $\rho_{\rm Heads}$, $\rho_{\rm Tails}$, in these approaches the work probability for  $\frac{1}{2}\rho_{\rm Heads} + \frac{1}{2}\rho_{\rm Tails}$ (tossing a fair coin and preparing $\rho_{\rm Heads}$ or  $\rho_{\rm Tails}$ accordingly) is \emph{not} the equal mixture of the work probability for  $\rho_{\rm Heads}$ and $\rho_{\rm Tails}$. These approaches require an interpretation for the lack of this property, which is normally a consequence of the standard probabilistic interpretation.
	\end{enumerate}
	Independently of the preferred alternative, given the proliferation of  definitions one needs to set  clear objectives for these investigations. It is not complex to come up with novel definitions of quantum fluctuations, leading to  suitably complicated fluctuation theorems which differ from the standard ones due to the presence of quantum coherence. To make a proposal relevant, however, one should attempt to solve three main problems:  
	\begin{enumerate}
		\item[(a)] Give an answer to the above-mentioned conceptual questions, providing a thermodynamic interpretation of the failure of properties expected in the statistical interpretation and showing how these are recovered in the classical limit.
		\item[(b)] Find a definition of fluctuations leading to a notion of `quantum signature' that can be identified with a clear-cut notion of nonclassicality. In other words, one should define a precise notion of classicality and show that the identified quantum signature is an \emph{operational} feature that cannot be reproduced by \emph{any} model satisfying that classicality property.
		\item[(c)] Show that such quantum signatures are \emph{experimentally accessible} without requiring a very large amount of control (like knowledge of the initial state or the dynamics). Ideally witnessing quantum signatures should necessitate no more control than that required by the original TPM scheme.
	\end{enumerate}
What we set out to do in this work is to show that all these problems can be solved simultaneously in the context of heat fluctuations.

\section{The problem of quantum heat fluctuations}
\label{sec:problem}
 The impasse in the definition of heat and work fluctuations can be traced back to the information-disturbance tradeoff of quantum mechanics. Imagine the cold and hot systems, C and H, initially described by a density operator $\rho_{\Co \Ho }$, evolve under a closed dynamics described by a unitary $U$, $\rho_{\Co \Ho }(\tau) = U \rho_{\Co \Ho} U^\dag$. The average heat flow reads
\begin{equation}
\label{eq:weak}
Q := \tr{}{\rho_{\Co \Ho} H_C} - \tr{}{\rho_{\Co \Ho }(\tau) H_C}.
\end{equation}
In order to study fluctuations, classically we want to see this quantity as emerging from a microscopic process where energy $q$ is exchanged with probability $p(q)$ such that $Q=~\int dq p(q) q$.  In this case, the system  C starts at some phase space point $z_0$ with probability $p(z_0)$ and evolves deterministically to $z_\tau$, in which case $q = E(z_0) - E(z_\tau)$, with $E(z)$ the energy at point $z$. Hence
\begin{equation}
p(q) = \int dz_0 p(z_0) \delta(q - (E(z_0) - E(z_\tau)))
\end{equation}
 Quantumly, however,  the initial energy  is generally not well-defined. Non-commutativity forbids the construction of a joint probability (linear in $\rho_{\Co \Ho }$) that correctly reproduces the energy distributions of the initial and final states \cite{perarnau2017no}.

This long-standing problem dates back many decades and the standard way to overcome it has been to define fluctuations by a two-point-measurement (TPM) scheme \cite{campisi2011colloquium, esposito2009nonequilibrium}. The TPM scheme involves (1) projectively measuring the energy of C at the start, (2) letting the system dynamics unfold, and (3) projectively measuring the energy of C again at the end.  $p^{\rm TPM}(q)$ is then defined as the probability of recording an energy difference between the outcomes of the two projective measurements equal to $q$. 
 However, in this scheme $Q^{
\rm TPM} := \int dq  p^{\rm TPM}(q) q \neq Q$, since effects due to non-commutativity are expunged by the initial invasive measurement. The widely used TPM method has the advantage that, by construction, classical fluctuation theorems hold for $p^{\rm TPM}(q)$. The drawback is that we are effectively describing the heat fluctuations of a state disturbed (decohered) by the initial energy measurement, rather than those of the original state $\rho_{\Co \Ho }$. Quantum effects such as strong heat backflows provably never appear. Hence this solution cannot be satisfactory in the fully quantum regime, when $\rho_{\Co \Ho}$ includes energetic coherence and quantum correlations.

\subsection*{A different solution: quasiprobabilities} We argue that the root cause of the present difficulties is the attempt to give a classical stochastic account of quantum heat fluctuations. 
Since quantum mechanics does not allow for the construction of joint probabilities  of noncommuting quantities, a natural framework is to describe them by means of quasiprobabilities~\cite{hofer2017quasi}, i.e. real-valued functions $p(q)$ that sum to $1$ but are allowed to take values outside $[0,1]$. While relatively uncommon in the thermodynamic setting (but see e.g. \cite{allahverdyan2014nonequilibrium, solinas2015full, solinas2016, xu2018effects, lostaglio2018quantum}), the use of quasiprobabilities to probe quantum effects is widespread in other fields of quantum sciences, including quantum transport \cite{belzig2001full,bednorz2010quasiprobabilistic,hovhannisyan2019quantum},  quantum optics~\cite{bucca, wigner1932quantum}, quantum computing \cite{pashayan2015estimating,seddon2020quantifying} and condensed matter \cite{halpern2017jarzynski, alonso2019out}.
 Quasiprobabilities originated with the attempt to give a phase space description to quantum mechanics despite non-commutativity of position and momentum. The most famous example is the Wigner function \cite{wigner1932quantum}, which associates a quantum state with a quasiprobability distribution over positions and momenta whose marginals correctly reproduce the statistics collected in the corresponding quantum measurements \footnote{Interestingly, Wigner originally introduced his function to compute quantum corrections to thermodynamic quantities.}. 
	
The analogy to our scenario is that, in order to recover Eq.~\eqref{eq:weak}, we wish to reproduce the correct quantum statistics of energy measurements performed on the initial $\rho_{\Co \Ho }$ and final $\rho_{\Co \Ho }(\tau)$ states.
Formally, if $\pwm\itf$ denotes the quasiprobability of starting with energies $(E^{\rm C}_{i_\Co}, E^{\rm H}_{i_\Ho})$ before the process $U$ and ending up in the final energies  $(E^{\rm C}_{f_\Co}, E^{\rm H}_{f_\Ho})$ afterwards, then we wish
\begin{equation}
\nonumber
\sum_{f_\Co, f_\Ho} \pwm\itf = \tr{}{\rho_{\Co \Ho} \Pi^{i_{\Co} i_{\Ho}}}:=  p^{i_{\Co} i_{\Ho}}, 
\end{equation}
 \begin{equation}
\label{eq:marginalproperty}
\sum_{i_\Co, i_\Ho} \pwm\itf = \tr{}{\rho_{\Co \Ho}(\tau) \Pi^{f_{\Co} f_{\Ho}}}:=  p^{f_{\Co} f_{\Ho}}(\tau), 
\end{equation}
where \mbox{$\Pi^{i_{\Co} i_{\Ho}} \equiv \Pi^{i_{\Co}} \otimes \Pi^{i_{\Ho}}$} and $\Pi^{i_{\rm X}}$ are energy projection operators such that $H_{\rm X} = \sum_{i_{\rm X}} E^{\rm X}_{i_{\rm X}} \Pi^{i_{\rm X}}$ (X = C, H). Also note that $p^{i_{\Co} i_{\Ho}}$ ($p^{i_{\Co} i_{\Ho}}(\tau)$) denotes the statistics of an energy measurements carried out at the beginning (end) of the experiment.  
The condition in Eq.~\eqref{eq:marginalproperty} ensures that the fluctuations  $\pwm\itf$ reproduce the average heat flows of Eq.~\eqref{eq:weak}:
\begin{equation}
\label{QMargenau} \sum_{i_\Co, i_\Ho, f_\Co, f_\Ho} \pwm\itf (E_{i_\Co} - E_{f_{\Co}}) = Q.
\end{equation}
 \section{A quasiprobability for quantum heat fluctuations} 
\label{sec:quasi} 
 How do we choose a quasiprobability distribution among the many possible candidates? The analogy with quantum optics suggests different choices will be relevant in different scenarios. Nevertheless, one candidate stands out.  Recall that the TPM distribution reads 
 	\begin{equation}
 	\ptpm\itf := \tr{}{ \Pi^{f_\Co f_\Ho}(\tau)   \Pi^{i_\Co i_\Ho} \mathcal{D} \rho_{\Co \Ho}},
 	\end{equation}
 	where $ \Pi^{f_\Co f_\Ho}(\tau) :=  U^\dag  \Pi^{f_\Co f_\Ho} U$ and $\mathcal{D}$ denotes dephasing, the operation that removes all off-diagonal elements in the energy basis. Moreover, the TPM characteristic function is associated with a two-time correlation function \emph{computed on a decohered version} $\mathcal{D}\rho_{\Co \Ho}$ \emph{of the initial state} $\rho_{\Co \Ho}$ \cite{campisi2011colloquium}. This naturally suggests reintroducing coherence and quantum correlations by removing such dephasing:
 \begin{equation}
\label{eq:weakquasiprobability}
\pwm\itf := {\rm Re} \, \tr{}{ \Pi^{f_\Co f_\Ho}(\tau)   \Pi^{i_\Co i_\Ho} \rho_{\Co \Ho}},
\end{equation}
where we focus here on the real part (the imaginary part may be of independent interest \cite{kunjwal2018anomalous}). This is our proposal for a quasiprobability distribution for heat fluctuations in the quantum regime. 

Since every $\rho_{\Co \Ho}$ can be additively decomposed as  a term $\mathcal{D} \rho_{\Co \Ho}$ (block-)diagonal in the energy basis plus energetic coherences $\chi_{\Co \Ho}$,  i.e. $\rho_{\Co \Ho} = \mathcal{D} \rho_{\Co \Ho} + \chi_{\Co \Ho}$,  corrections to the standard scheme emerge only if \mbox{$\chi_{\Co \Ho} \neq 0$}:
\begin{equation*}
\pwm\itf = \ptpm\itf + {\rm Re} \, \tr{}{ \Pi^{f_\Co f_\Ho}(\tau)   \Pi^{i_\Co i_\Ho} \chi_{\Co \Ho}}.
\end{equation*}
The quantum correction to $\ptpm\itf$ is bounded by $\| \chi_{\Co \Ho}\|$, which is a coherence measure \cite{aberg2006superposition, marvian2016how}.
 $\pwm$ satisfies the marginal properties of Eq.~\eqref{eq:marginalproperty} and hence it reproduces the quantum heat flow of Eq.~\eqref{QMargenau}. As expected,  when $\chi_{\Co \Ho} \neq 0$ in general $\pwm$ does not describe a stochastic process, since it can be negative: in fact $\pwm\itf~\in~[-1/8,1]$\footnote{Ref.~\cite{allahverdyan2014nonequilibrium} (Eq. 41) or  \cite{zhang2011matrix} (Theor.~7.5).}.  It is worth noting that $\pwm$ recovers the definition of a quasiprobability distribution for two noncommuting observables 
given by Margenau and Hill in 1961~\cite{margenau1961correlation} and known since the late 1930s \cite{terletsky1937limiting}. 
It also takes the same form as the work quasiprobability proposed by Allahverdyan in Ref.~\cite{allahverdyan2014nonequilibrium},  whose negativities were however dismissed as an unphysical feature. Here we present a different picture.  

Firstly, while $\pwm$ can show negativities, it can be experimentally accessed in a probing scheme that naturally extends the TPM scheme 
(see App.~\ref{app:A} for more details). To estimate $\pwm$ one has to replace the first  of the two projective measurements of the TPM scheme with a \emph{weak measurement},  which is achieved by very weakly coupling the system to either a continuous variable pointer or a qubit probe  which is then projectively measured. Appealingly, $\pwm$ can hence be reconstructed in a `minimally disturbing' version of the TPM scheme. 

 Secondly, negativity is a precise signature of nonclassicality. In fact, $\pwm\itf$ is proportional to a quantity inferred from weak measurements and known as the (generalized) \emph{weak value}~\cite{dressel2015weak}, first introduced by Aharonov \emph{et.al.} \cite{aharonov1988result,dressel2014colloquium} and later generalized to mixed states \cite{wiseman2002weak, dressel2010contextual}. It then follows from the extensions \cite{lostaglio2018quantum, kunjwal2018anomalous} of Pusey's theorem \cite{pusey2014anomalous} that the negativities of $p^{\rm W}$ are witnesses of a strong form of nonclassicality known as generalised contextuality \cite{spekkens2005contextuality}. Specifically, the statistics collected by the weak measurement scheme probing $p^{\rm W}$ cannot be explained by \emph{any} noncontextual hidden variable model, a claim valid even in the presence of noise~\cite{kunjwal2018anomalous}.  This is directly related to the absence of \emph{any} classical stochastic process explaining the relevant statistics~\cite{spekkens2008negativity}.

Thus far we showed that $\pwm$ is a natural extension of the TPM scheme, both formally and experimentally; that it reproduces the heat flows of $Q$, including strong heat backflows due to initial entanglement; and that its negativity is a precise signature of strong nonclassicality.
However, what is the \emph{thermodynamic} significance of negativity and how does it impact the observed heat flows? We now answer these questions.

\subsection{Thermodynamic role of negativities in the heat fluctuations}  
\begin{figure*}
\centering
\subfloat{\includegraphics[width=0.48\linewidth]{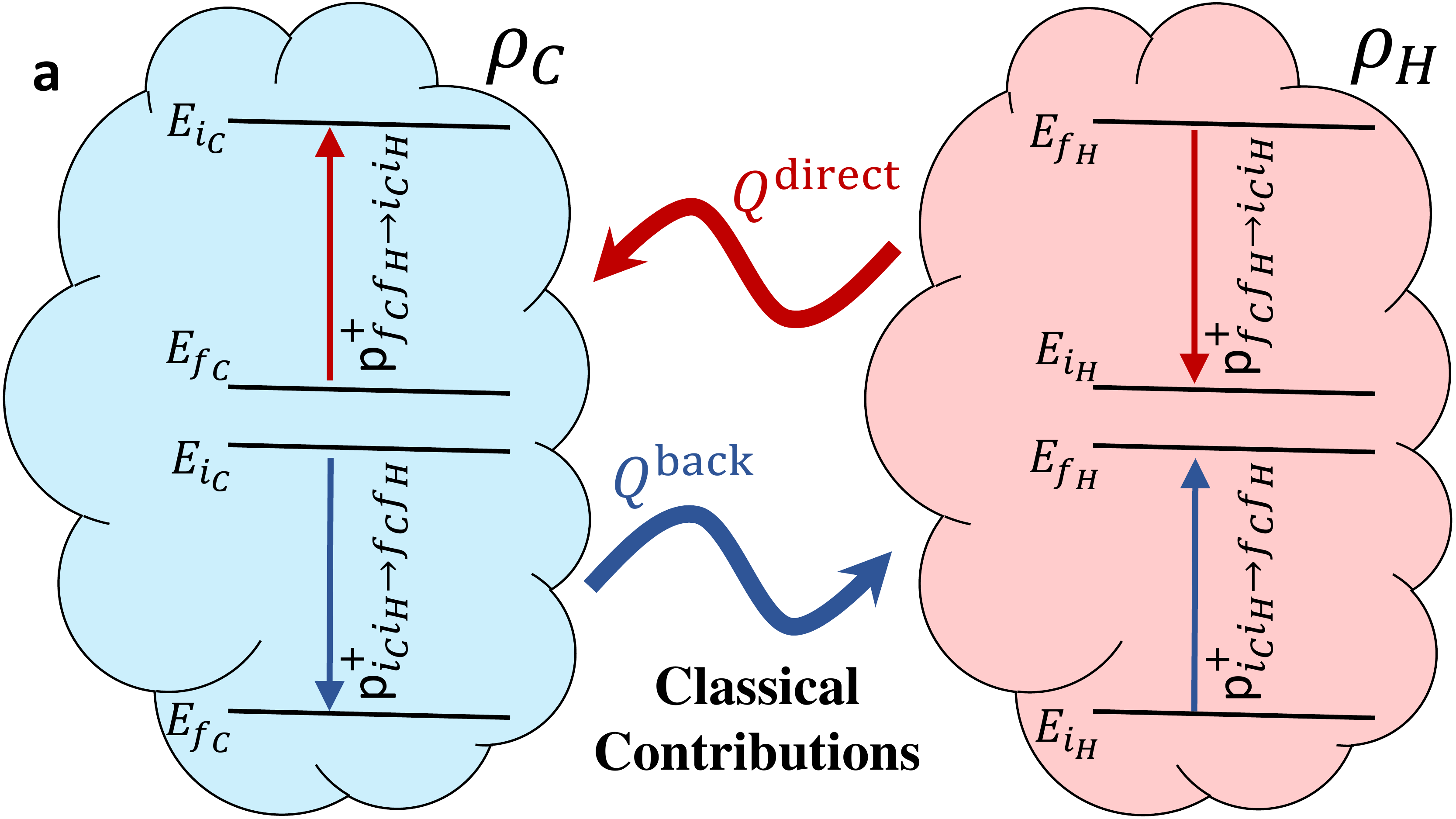}}
\qquad
\subfloat{\includegraphics[width=0.48\linewidth]{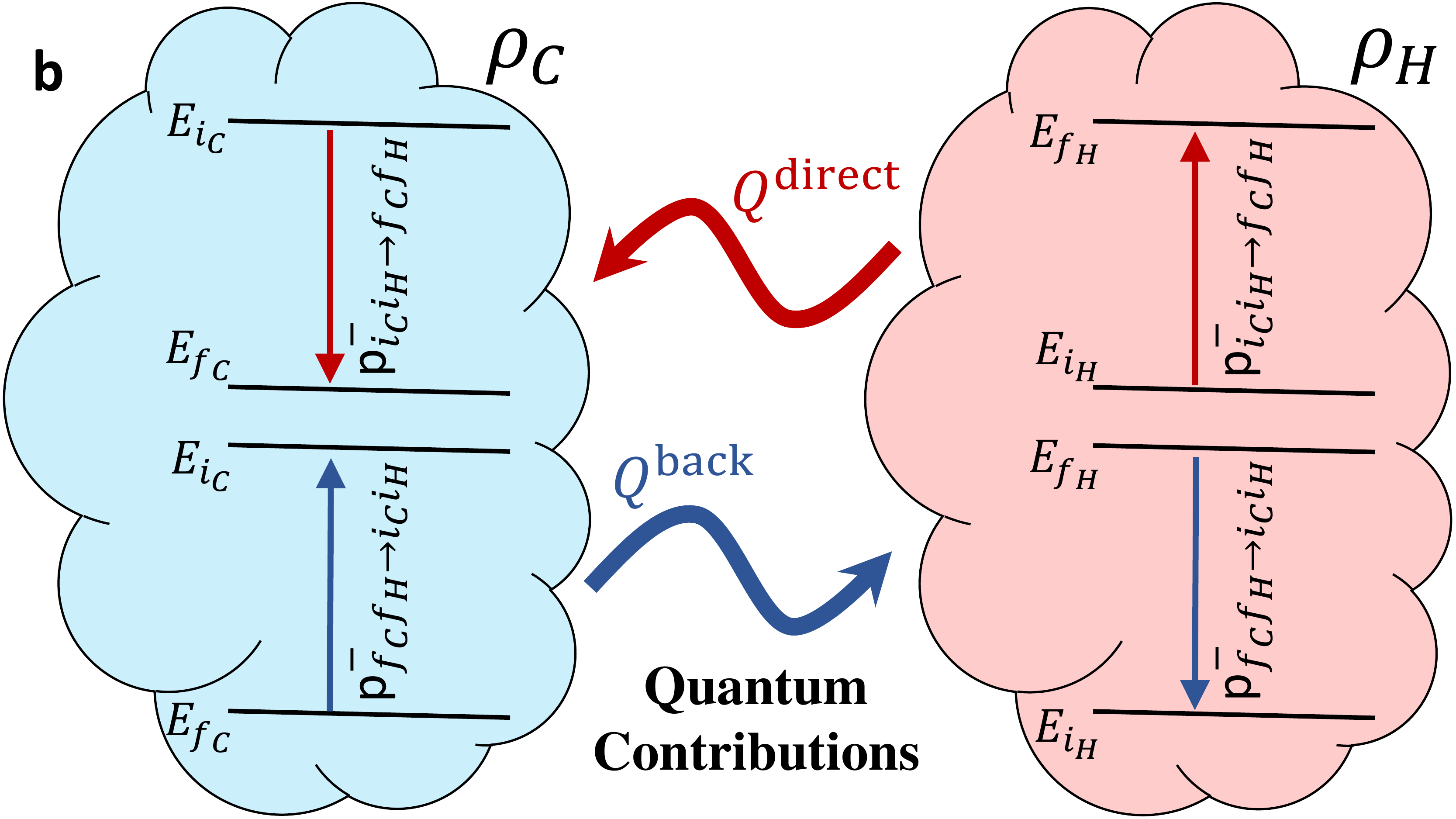}}
	\caption{ Schematic illustration of the classical and quantum contributions to the overall heat flow $Q$ between a cold (C) and a hot (H) system. $Q$ is an average of direct (H $\rightarrow$ C) and back flows (C $\rightarrow$ H). {\bf a}, Classical contributions (positive probability events) corresponds to standard contributions, where the loss of energy from H contributes to the flow towards C and vice versa. {\bf b}, Quantum contributions (negative probability events) effectively act as contributions in which the loss of energy from H contributes to the flow (C $\rightarrow$ H).} 
	\label{fig:heatflows}
\end{figure*}

 To single out the meaning of negativities, let us split the quasiprobability in positive and negative components: $ \pwm\itf = p^+_{i_\Co i_\Ho \rightarrow f_\Co f_\Ho} + p^-_{i_\Co i_\Ho \rightarrow f_\Co f_\Ho}$, where $p^\pm_{i_\Co i_\Ho \rightarrow f_\Co f_\Ho}=  \pwm\itf$ if $ \pwm\itf$ is positive (negative), and zero otherwise. 
Recall that, with our sign convention, $Q>0$ means backflow. Then, the net flow $Q$ can be decomposed into direct (H $\rightarrow$ C) and back (C~$\rightarrow$~ H) flows:
\begin{equation}
Q = Q^{\rm back} - Q^{\rm direct},
\end{equation}
 where the `back' term includes all contributions to the flow $\Co \rightarrow \Ho$ and the `direct' term includes all contributions $\Ho \rightarrow \Co$: 
\begin{align}
Q^{\rm back} = \sum_{E_{i_\Co}>E_{f_{\Co}}} (p^+_{i_\Co i_\Ho \veryshortarrow f_\Co f_\Ho} - p^-_{ f_\Co f_\Ho \veryshortarrow i_\Co i_\Ho }) \Delta E_{i_\Co f_{\Co}},
\label{Qback} \\
Q^{\rm direct} = \sum_{E_{i_\Co}>E_{f_{\Co}}} (p^+_{f_\Co f_\Ho \veryshortarrow i_\Co i_\Ho } - p^-_{ i_\Co i_\Ho \veryshortarrow f_\Co f_\Ho })  \Delta E_{i_\Co f_{\Co}}.
\label{Qdirect} 
\end{align}
These equations give a very suggestive interpretation of the role of negative probabilities in heat flows (see Fig.~\ref{fig:heatflows}). Since $E_{i_\Co} > E_{f_{\Co}}$, $Q^{\rm back}$ has two positive contributions: 
1.~From the transition $i_{\Co} \rightarrow f_{\Co}$, which removes energy from C (as expected).
2.~From the transition $f_{\Co} \rightarrow i_{\Co}$ \emph{when}  $p^-_{ f_\Co f_\Ho \veryshortarrow i_\Co i_\Ho }<0$.  Classically, the transition $f_{\Co} \rightarrow i_{\Co}$ would add energy to C, yet it contributes to the backflow from C to H when the correspondent quasiprobability turns negative. 
Symmetrically, a transition taking energy away from C can contribute to the direct-flow, when $p^-_{ i_\Co i_\Ho \rightarrow f_\Co f_\Ho }<0$. These counterintuitive heat flow contributions signal the breakdown of a stochastic description and the onset of quantum effects,  as depicted in Fig.~\ref{fig:heatflows}.

 In particular, we can understand strong flows as the result of negative probabilities, which are destroyed by the measurement stage of the TPM scheme  (a projective energy measurement on C and H forces $\pwm>0$). Furthermore, Eq.~\eqref{Qdirect} suggests  \textit{negativities may lead to strong direct-flows as well}. 
 
 It is important to bear in mind that negativity will not always result in strong heat flows.
Since only $Q =Q^{\rm back} - Q^{\rm direct}$ is observed, negativites from $Q^{\rm back}$ and $Q^{\rm direct}$ must not cancel each other if we are going to observe an overall effect in the heat flow. 

\subsection{Classical limit}

 When discussing quantum effects in heat or work fluctuations, it is important to distinguish between two different forms:
	\begin{enumerate}
		\item Quantumness due to  correlations or entanglement in the initial state, such as that responsible for strong heat backflows.
		\item Quantumness due to the superpositions created by $U$ \emph{during} the dynamics.
	\end{enumerate}

The first form of nonclassicality is the one preventing a straighforward description of quantum heat exchanges by means of a classical stochastic process. Classicality in the first sense  emerges in the expected regimes. The typical mechanism is decoherence. If $\Co$ and $\Ho$ are not protected from the external environment, superpositions of different energy states are quickly suppressed ($\chi_{\Co \Ho} \rightarrow 0$) and so $\pwm~\rightarrow ~\ptpm \geq 0$, i.e. negativity disappears. 
 Decoherence may be also due to the measurement process, which is the mechanism that introduces classicality in the TPM scheme.
 
  Another mechanism relevant at mesoscopic scales is coarse-graining. Suppose one can only access the heat exchanges between two clusters, each containing $N$ particles, but cannot keep track of the process at the level of the individual constituents. We idealise the situation via an average Hamiltonian $H_{\Co \Ho} \approx \sum_{i=1}^N (H_{\Co_i} + H_{\Ho_i})/N$ and a dynamics $U \approx \otimes_{i=1}^N U_{\Co_i \Ho_i}$.
	 Then, whatever the initial state $\rho_{\Co \Ho }$ is, $\pwm \approx \ptpm + O(1/N)$ \footnote{This follows from the fact that $[U^\dag H_{\Co \Ho}U, H_{\Co \Ho}] \propto 1/N$.}.
	 The ratio between fluctuations and average flows, on the other hand, is suppressed as $O(1/\sqrt{N})$. There is hence an intermediate, ``mesoscopic'' regime in which quantum effects are irrelevant but fluctuations are not yet negligible. 
	 
	  The second form of nonclassicality was studied in \cite{lloyd2015no}. In this study it was shown that for initially uncorrelated bipartite system energy cannot flow without generation of quantum correlations in the form of discord during the unitary process. This form of quantumness is the one responsible for the differences between the classical stochastic process predicted by the TPM scheme and the one obtained within a fully classical description (e.g., a classical mechanics model), whenever the latter is well-defined. Note that, since we recover the TPM probability in the regimes discussed above, we can exploit previous results on the emergence of this second form of classicality when the superpositions created by $U$ \emph{during} the dynamics can be neglected~\cite{jarzynski2015quantum}. 
	 
	  We conclude that our definition of heat fluctuations appears to provide a clear path to the emergence of classicality, from negative quasiprobability all the way up to a classical mechanical model, passing through the standard TPM scheme that plays the role of a ``quasi-classical'' description.

\section{Heat flows witnessing negativities in a two qubit system} 
\label{sec:qubits}
 By attempting to derive a  stochastic interpretation of quantum fluctuations, we arrived at a picture in which quantum effects are incorporated as non-classical features in the heat exchange, such as contributions to the energy flow from C to H associated with transitions to \emph{higher} levels in C. From a practical perspective, however, the next step is to ask if these non-classical features can be directly related to observable effects and if these effects can be accessed in realistic experiments. We answer both questions in the positive.

 We start with the archetypal (and fully solvable)  idealized two-qubit scenario. 
Since $U$ injects no work, nontrivial energy exchanges can only happen if the two qubits are resonant.
\begin{theorem} \label{th:qubitspecificbounds} 
Let $\rho_{\Co \Ho}$ be a two qubit system with thermal marginals ($\beta_{\Co} \neq \beta_{\Ho}$), $H_\Co = H_\Ho = E \ketbra{1}{1}$ and $U$ a unitary injecting no work, $[U, H_\Co + H_\Ho] = 0$. If $\pwm$ is nonnegative,
	\begin{equation}
	\label{eq:qubitbound}
	|Q| \leq \frac{2 + e^{\beta_\Ho E} + e^{\beta_\Co E} }{e^{\beta_{\Co}E} - e^{\beta_{\Ho}E}} |\qtpm|.
	\end{equation}
	\end{theorem}
See App.~\ref{subsec:proof1} for the proof.  This `heat flow inequality' shows that strong enough direct or backflows of heat can only occur when negativity is at play in $\pwm$. Note that to violate the bound of Eq.~\eqref{eq:qubitbound}, we necessarily  witness $|Q| > |\qtpm|$, i.e. a direct or inverse flow bigger than the corresponding flow observed in the TPM scheme. While violations of Eq.~\eqref{eq:uncorrelatedbound},  reported in Ref.~\cite{micadei2019reversing}, can occur in a purely classical setup, due to initial correlations in the energetic degrees of freedom, violations of Inequality~\ref{th:qubitspecificbounds} imply negativity, which is a proxy for genuinely quantum effects. 

\subsection{Experimental verification}

Next, we analyze the violations of Inequality \ref{th:qubitspecificbounds}. In  recent NMR experimental setups~\cite{micadei2019reversing,pal2018experimental}, the heat backflow between two local-thermal qubits with initial correlations was measured. Here we focus on the data of Ref.~\cite{micadei2019reversing} and
 show that, while other general approaches (e.g., Ref.~\cite{jennings2010entanglement}) fail to witness quantum effects for this setup,
Inequality~\ref{th:qubitspecificbounds} is violated.

 In the experiment, the heat backflow $Q$ between two nuclear spin $1/2$, prepared in an initial locally thermal state, was measured. We denote by $\sigma_i$, $i=x,y,z$, the Pauli matrices. The initial Hamiltonian is $H_{\Ho \Co} = H_{\Ho} + H_{\Co}$ with $H_{\Co(\Ho)} = h \nu (1- \sigma^{\Co(\Ho)}_z)$  and $h$ Planck's constant.  A locally thermal state is prepared with the form
\beq
\label{eq:stateexperiment}
\rho_{\Ho \Co}=\rho_{\Ho}\otimes\rho_{\Co}+\gamma\ketbra{01}{10}+\gamma^{*}\ketbra{10}{01},
\eeq 
where $\rho_{\Ho(\Co)} = e^{-\beta_{\Ho(\Co)} H_{\Ho(\Co)}}/ \tr{}{e^{-\beta_{\Ho(\Co)} H_{\Ho(\Co)}}} $. The unitless rescaled inverse temperatures  reported in the experiment are  $\beta_{\Ho}=0.9618$ and $\beta_{\Co}=1.13$, with the gap $\nu=1{\rm kHz}$. In the correlated scenario, $\gamma=-0.19$, whereas in the uncorrelated scenario  $\gamma=0$.  

During the process, the nuclear spins are coupled via the interaction Hamiltonian \mbox{$H_{\rm int}=J\pi \hbar/2(\sigma_x^{\Ho}\sigma_y^{\Co}-\sigma_y^{\Ho}\sigma_x^{\Co})$} with $J=215.1{\rm Hz}$. Since $[H_{\rm int}, H_{\Ho \Co}] = 0$, this generates an energy-preserving unitary. A simple yet important observation is the following: the measurement of the heat exchange when $\gamma = 0$ corresponds to the measurement of $Q^{\rm TPM}$ for the case $\gamma = - 0.19$. This follows from the fact that the initial energy measurement on $\Co$ and $\Ho$ (the first step of the TPM protocol) has the effect of setting $\gamma= 0$. Thanks to this observation, the data collected in the experiment suffices to test Inequality~\ref{th:qubitspecificbounds}.

\begin{figure}[t!]
	\includegraphics[width=0.85\linewidth]{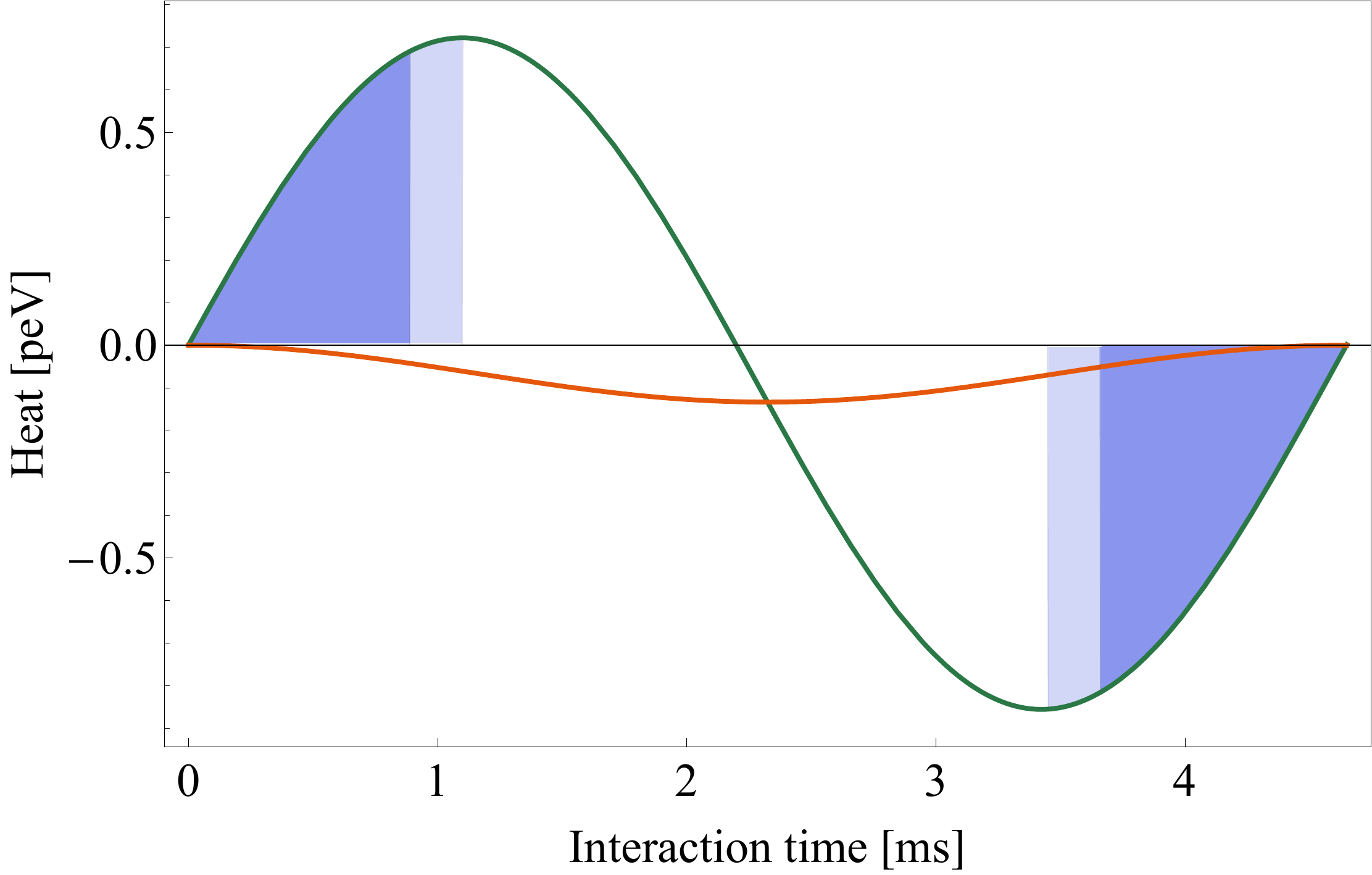}
	\caption{ Heat transfer $Q$ and $Q^{\rm TPM}$  (green and orange curves, respectively) for the experiment in Ref.~\cite{micadei2019reversing}. The shaded areas indicate the regime of negative probabilities. The dark blue shaded area indicates the violation of Inequality~\ref{th:qubitspecificbounds}. }  
	\label{fig:qubitslutz}
\end{figure}

In Fig.~\ref{fig:qubitslutz} we plot the heat transfer between qubits $\Co$ and $\Ho$ for various interaction times. The green line corresponds to the heat flow when the qubits are initially quantum correlated, whereas the orange line corresponds to initially uncorrelated qubits. Equivalently, this corresponds to the measurements of $Q$ (green) and $Q^{\rm TPM}$ (orange).  $Q>0$ corresponds to backflow from $\Co$ to $\Ho$, whereas $Q<0$ corresponds to direct flow from $\Ho$ to $\Co$.
The shaded areas indicate the interaction times in which negativity in heat transfer quasiprobability  is present.
The dark blue shaded area indicates the interaction times for which Inequality~\ref{th:qubitspecificbounds} successfully witnesses negativity for the relevant experimental parameters.

\begin{figure}[t!]
\includegraphics[width=0.8\linewidth]{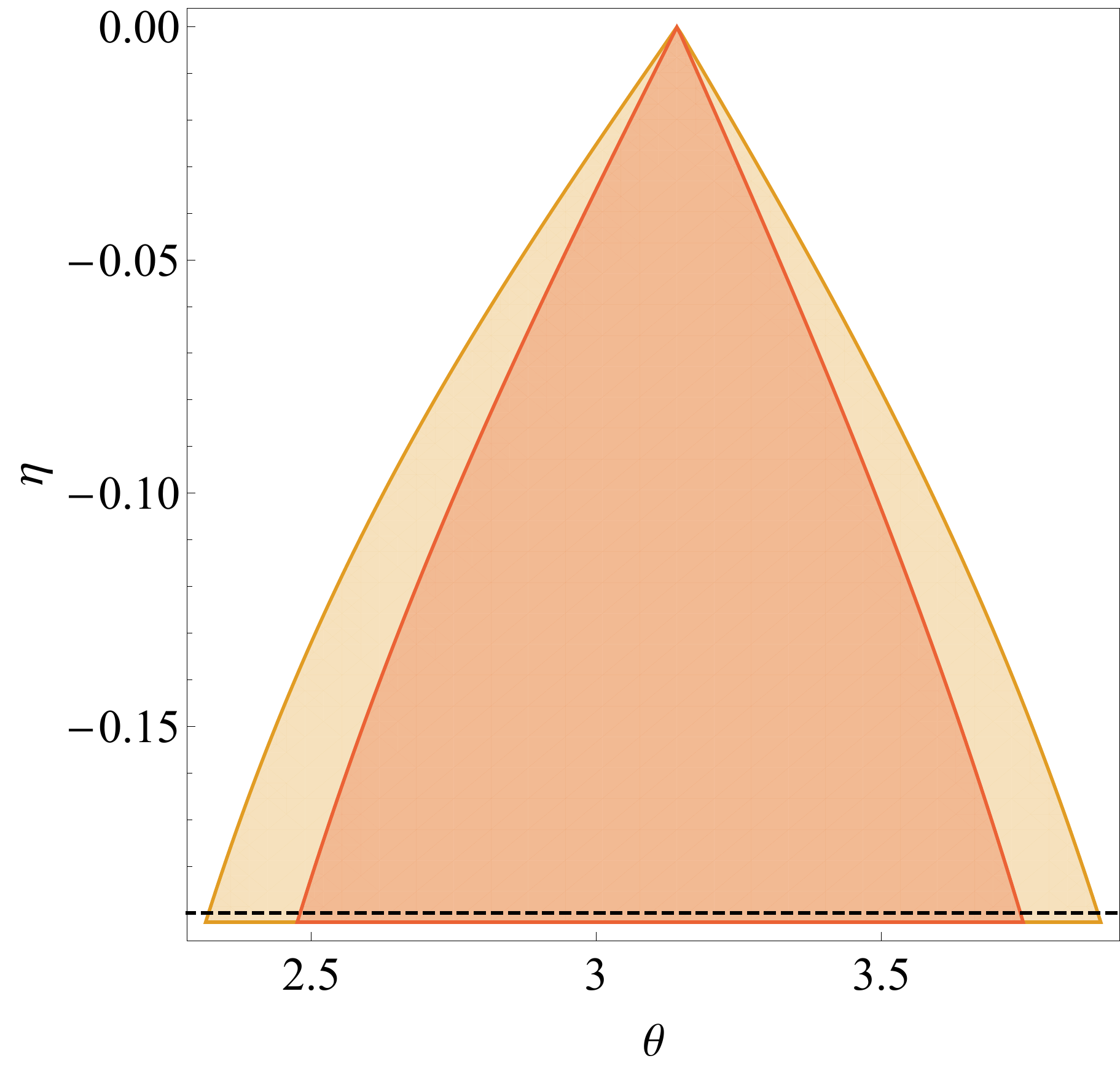}
	\caption{ 
	Region of negative probabilities (large  trianguloid) and violations of Inequality~\ref{th:qubitspecificbounds} (darker trianguloid) in the parameter space $(\theta,\eta)$. $\theta$ characterizes the unitary evolution (the rotation in the $\ket{01}$, $\ket{10}$ subspace) and $\eta$ the initial energetic superposition between $\ket{01}$, $\ket{10}$. 
In App.~\ref{app:B} we show that for fixed $\beta_{\Co}$, $\beta_{\Ho}$, and a simple reparametrization, the parameter space is restricted to $(\theta, \eta,P_{00})$, with 
$P_{00}$ the population in the state $\ket{00}$. 
In the experiment \cite{micadei2019reversing},  $\beta_{\Co}=1.13$,  $\beta_{\Ho}=0.962$, $P_{00} = 0.547$, and $\eta=-0.19$, which is indicated by the dashed line at the bottom of the figure. } 
	\label{fig:qubitbounds}
\end{figure}

While in the experiment only the backflow regime was explored, which is sufficient for detecting negativity, here we theoretically extend the interaction time to the regime where direct flow is observed.  We see that also in this regime, the measurement of the heat flows would allow us to detect the negativity of the quasiprobability,  showing that quantum effects are not only present in backflows.
We further extend the investigation of Inequality~\ref{th:qubitspecificbounds} beyond the specific parameters of Ref.~\cite{micadei2019reversing}, see Fig.~\ref{fig:qubitbounds}.   
We conclude that negativity can be detected for a wide range of parameters in accessible experiments.

 Using the experimental parameters of Ref.~\cite{micadei2019reversing}, the smallest eigenvalue of the partial transpose $\rho_{\Co \Ho}^{T_\Ho}$ of $\rho_{\Co \Ho}$ is $\approx 0.0014$, hence $\rho^{ T_{\Ho}}_{\Co \Ho} \geq 0$. By the Peres-Horodecki criterion, $\rho_{\Co \Ho}$ is separable.  This shows that the heat fluctuations can have negativities, as detected by Inequality~\ref{th:qubitspecificbounds}, even when $\rho_{\Co \Ho}$ is not entangled.

  We mention in passing a nontrivial result that holds specifically for two qubit systems: \textit{the existence of quantum correlations in the energy degrees of freedom is necessary for witnessing a backflow for an appropriate energy-preserving dynamics} (see App.~\ref{app:B}).  This is the reason the orange curve in Fig.~\ref{fig:qubitslutz} is all below the $x$-axis.

\subsection{Nonideal heat exchange}  

While it is commonplace to assume that $U$ injects an amount of work into the system that can be neglected in calculations, strictly speaking Inequality~\ref{th:qubitspecificbounds} holds only in an ideal case. Next, we extend Inequality~\ref{th:qubitspecificbounds} to realistic heat exchange scenarios, in which (1) the energy levels are not resonant, (2) the dynamics injects some work:

\begin{theorem} \label{th:qubitspecificboundsnonideal} 
	Let $\rho_{\Co \Ho}$ be a two qubit system with thermal marginals ($\beta_{\Co} \neq \beta_{\Ho}$), $H_\Co = E_C \ketbra{1}{1}$, $H_\Ho = E_H \ketbra{1}{1}$ and U a unitary dynamics. Suppose
	\begin{enumerate}
		\item $U$ is close to an energy preserving unitary $\tilde{U}$: \mbox{$
			\| U - \tilde{U} \| \leq \epsilon$}. 
		\item The gap is not too wide: $\Delta := \frac{|E_{\Co} - E_{\Ho}|}{2 \bar{E} }< \frac{1-R}{1+R}$, where  $\bar{E}:= (E_{\Co} + E_{\Ho})/2$ is the average energy and   
		\mbox{$R = (1+ e^{\beta_{\Ho} E_{\Ho}})/ (1+ e^{\beta_{\Co} E_{\Co}})$}.
		\item The energy flow is large enough that we can distinguish a heat contribution: $|Q| > 2 \epsilon \bar{E}$.
	\end{enumerate}
	If $\pwm$ is nonnegative,
	\begin{small}
		\begin{equation}
		\label{eq:qubitboundnonideal}
		|Q| \leq \frac{1+ R + \Delta (1+R)}{1- R - \Delta (1-R)} |\qtpm| + 4 \epsilon \bar{E} \frac{2 + \Delta (1+R)}{1-R - \Delta(1-R)} .
		\end{equation}
	\end{small}
\end{theorem}
When $\epsilon \rightarrow 0$, $\Delta \rightarrow 0$ we recover Inequality~\ref{th:qubitspecificbounds}. The above inequality is suitable to be applied in realistic experimental scenarios. Note that the condition of unitarity is not crucial and Inequality~\ref{th:qubitspecificboundsnonideal} can be extended to quantum channels. The proof of this inequality, stronger bounds, and a case study are presented in App.~\ref{app:nonidealflow}.  This shows that analogue heat flow inequalities can be derived beyond the ideal regime.

\section{A quantum exchange fluctuation theorem} 
\label{sec:qxft}
 The above heat flow inequality is limited to two-qubit systems. We now derive a heat flow inequality valid for arbitrary finite-dimensional systems. This is obtained through the derivation of a quantum exchange fluctuation theorem  (QXFT) in the spirit of \cite{allahverdyan2014nonequilibrium, kwon2019fluctuation}. 

Denote by  $\Delta I~:=~ I_{f_{\Co} f_{\Ho}}~-~I_{i_{\Co} i_{\Ho}}$ with
	\beq
	I_{i_{\Co} i_{\Ho}}=\log \frac{\tr{}{\Pi^{i_\Co i_\Ho} \rho_{\Co \Ho}} }{\tr{}{\Pi^{i_\Co} \rho_{\Co}} \tr{}{\Pi^{ i_\Ho} \rho_{\Ho}} },
	\eeq
	the elements of the classical mutual information. Assuming $E_{i_\Co} - E_{f_\Co} \approx E_{f_\Ho} - E_{i_\Ho}$ for all nonzero $\pwm\itf$, we obtain the XFT in the full quantum regime
\beq
\label{eq:XFT}
\ave{e^{\Delta I +\Delta\beta \Delta E_{ i_{\Co} f_{\Co}}}}_{ W}=1+\bar{\chi}.
\eeq 
The r.h.s of Eq.~(\ref{eq:XFT}) consists of 1, contribution from the populations, and $\bar{\chi}$, involving contributions from the coherence terms in the energy eigenbasis
\beq
\label{eq:chi}
\bar{\chi}=\sum_{l,k\neq m}\frac{\rho_{ll}}{\rho_{kk}}{\rm Re}\lbrace\rho_{km}\bra{l}U\ket{k}\bra{m}U^{\dagger}\ket{l}\rbrace,
\eeq 
which are absent in the TPM scheme. Note that 
\begin{enumerate}
	\item If $\rho_{\Co \Ho}$ is classically correlated in the energy basis, $\bar{\chi} =0$, recovering the main result of Ref.~\cite{jennings2012exchange}.
	\item If, furthermore, $\rho_{\Co \Ho}$ is uncorrelated, $\Delta I = 0$ and we recover Jarzynski's original result \cite{jarzynski2004classical}.
\end{enumerate}
Eq.~\eqref{eq:XFT} is thus the generalization of the XFT to the full quantum regime. The proof of the quantum XFT is provided in App.~\ref{app:C}.

\subsection{Witnessing negativity in arbitrary dimension}
Using the quantum XFT, we derive a general inequality witnessing negativity in any dimension: : 
\begin{theorem} \label{th:xftbounds} 
Let $\rho_{\Co \Ho}$ be an arbitrary finite-dimensional system with thermal marginals ($\beta_{\Co} \neq \beta_{\Ho}$) and $U$ a unitary that injects no work ($E_{i_\Co} - E_{f_\Co} = E_{f_\Ho} - E_{i_\Ho}$). If $\pwm$ is nonnegative,
	\begin{equation}
	\label{eq:xftbound}
	Q \leq -\frac{\ave{\Delta	I}_{\rm W}}{\Delta \beta}+\frac{\log(1+\bar{\chi})}{\Delta \beta}.
	\end{equation}
	\end{theorem}
This result can be derived by applying Jensen's inequality to Eq.~(\ref{eq:XFT}). $\ave{\Delta	I}_{W}$ in Eq.~(\ref{eq:xftbound}) is a classical mutual information-like term satisfying $\ave{\Delta	I}_{\rm W}=\ave{\Delta	I}_{\rm TPM}$ if the initial state only has classical correlations in the energy basis, and  $\ave{\Delta	I}_{W}=0$ in the absence of initial quantum and classical correlations \footnote{Note that the term in Eq.~(\ref{eq:chi}) may diverge as $\rho_{kk}\rightarrow 0$. In  App.~\ref{app:C} we derive a different bound for the heat in Eq.~(\ref{eq:xftbound}) that bypasses this problem. Also note that neither $\ave{\Delta	I}_{\rm W}$ nor $\ave{\Delta	I}_{T\rm PM}$ admit a simple interpretation as the change in classical mutual information in the energetic degrees of freedom.}. Violations of the inequality imply that some quasiprobabilities are necessarily negative. Note that the inequality becomes trivial if no initial coherence is present ($\chi_{\Co \Ho} =0$), as expected. In App.~\ref{app:C}  we provide a second heat flow inequality valid for general finite-dimensional systems.

For finite-dimensional systems with equal Hamiltonians and nondegenerate energy gaps, we  prove in App.~\ref{app:C} an additional generic property of negativity: 
 \textit{If all quasiprobabilities that contribute to direct or backflow are negative, then the flows are necessarily strong, i.e. $|Q|>|Q^{\rm TPM}|$}. 

\subsection{Nonideal heat exchange} 
Inequality~\ref{th:xftbounds} can be generalized to situations in which the injected work cannot be neglected. If for all nonzero $\pwm\itf$ one has $|E_{i_\Co} - E_{f_\Co} -( E_{f_\Ho} - E_{i_\Ho})| \leq \epsilon$, almost the same derivation allows us to generalise Eq.~\eqref{eq:xftbound} to 
\begin{equation}
p^{\rm W} \textrm{ nonnegative} \Rightarrow Q \leq -\frac{\ave{\Delta	I}_{\rm W}}{\Delta \beta}+\frac{\log(1+\bar{\chi})}{\Delta \beta} - \frac{\beta_{\Ho}\epsilon}{\Delta \beta}.
\end{equation}
A symmetric inequality can be derived for the energy flowing into $\Co$. 

 \subsection{Case study: heat exchange between two qutrits}
\label{subsec:case_study}
In this section we investigate the heat flows between two qutrits and show that Inequality~\ref{th:xftbounds} can witness negativity in these higher dimensional systems as well.
The Hamiltonian of the qutrits is assumed to take the form $H_{\Co} = H_{\Ho} = \sum_{n=0}^{2} E_n \ketbra{n}{n}$ with
  $E_0 = 0$, and  no degeneracy of the energy gaps (the `Bohr spectrum').
The state  $\rho_{\Co \Ho}=\sum_{i,j =0}^8 \rho_{ij}\ketbra{i}{j}$, can be expressed by the elements  $\rho_{ij}=\rho^{*}_{ji}=\eta_{ij}e^{-i\xi_{ij}}\sqrt{\rho_i \rho_j} \quad \forall \, i\neq j$,  the population $ \rho_{i} \equiv \rho_{ii}$, and the parameters  \mbox{$\eta_{ij} \in [0,1]$} and $\xi_{ij}\in \mathbb{R}$. Here, we used the natural labeling $(00,01,02,10,11,12,\dots) \equiv (0,1,2,3,4,5,\dots)$. 

\begin{figure}[t!]
\includegraphics[width=0.8\linewidth]{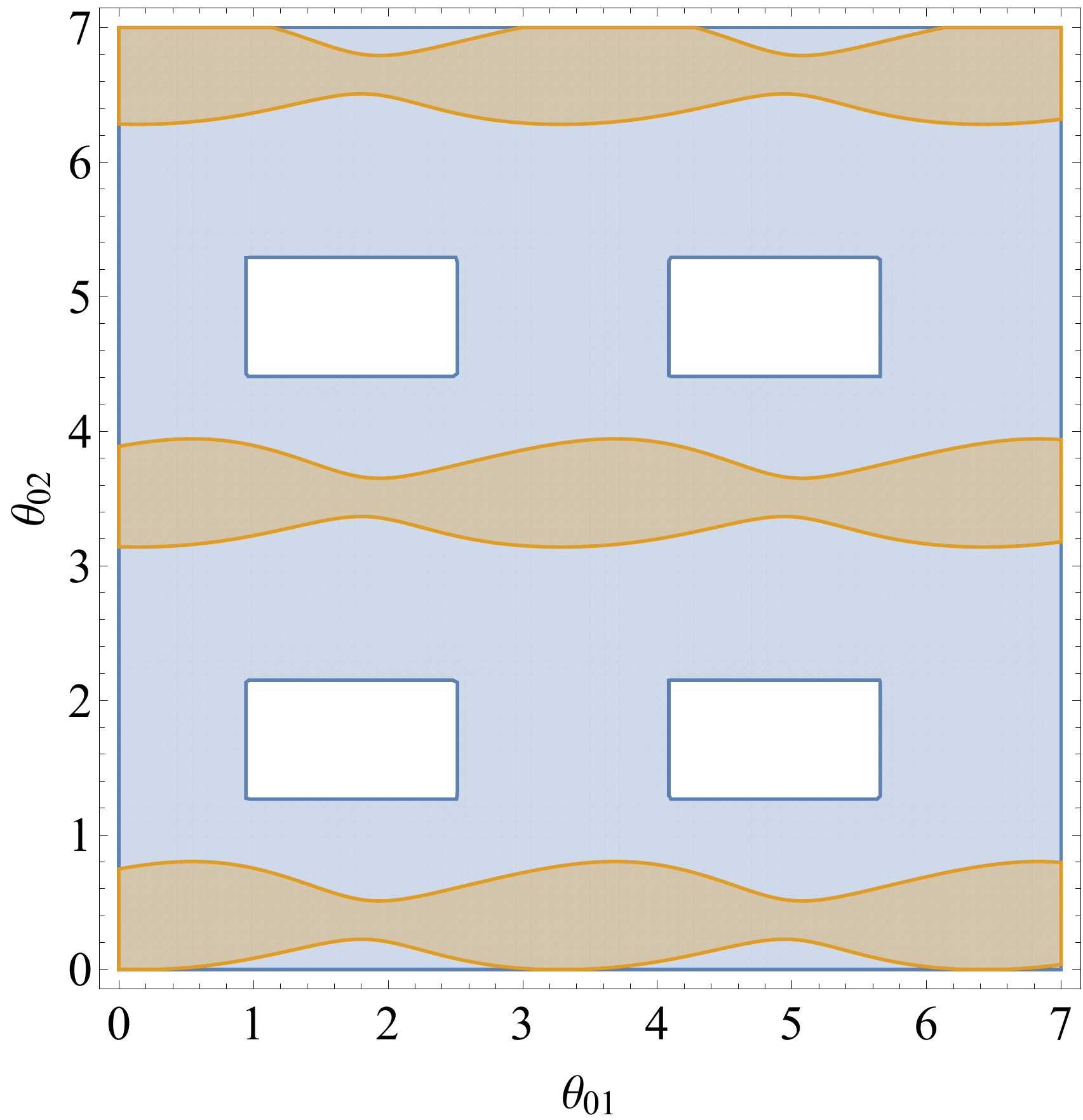}
	\caption{ 
 Negativity of the  quasiprobabilities (all shaded areas) and the violation of Inequality~\ref{thm:newbound} (yellow shaded area) for different interaction protocols determined by $\theta_{01} $ and $\theta_{02} $. Parameters: $\beta_{\Ho}=0.3$, $\beta_{\Co}=1.3$, $\theta_{12}=\theta_{02}$, $E_1=1$, $E_2=1.15$,  $\xi_{ij} = 0$, $\eta_{ij}= 1$ for $(i,j) = \{(1,3), (2,6), (5,7)  \}$, $\rho_0=0.3$, $\rho_5 =0.03$, $\rho_7 =0.07$, and $\rho_8 =0.06$.}  
	\label{fig:qutrits}
\end{figure}

 By imposing the constraint that the marginal states $\tr{\Co}{\rho_{\Co\Ho}}$ and $\tr{\Ho}{\rho_{\Co\Ho}}$  are thermal, and following the same reasoning as in the two qubit scenario, heat flows between two qutrits are determined by: the six parameters $\eta_{ij}$, $\xi_{ij}$ for $(i,j) = \{(01,10) \equiv (1,3), (02,20) \equiv (2,6) ,(12,21) \equiv (5,7)  \}$; and four undetermined populations $\rho_i$ that must comply with the nonnegativity of $\rho_{\Co\Ho}$.
 Furthermore, all possible heat flows achievable by energy preserving unitaries can be realised by $U = \oplus_{i=1}^3 U^{(i)}$, where $U^{(i)}$ are real $2$d rotations within the three above-mentioned manifolds, fixed by three angles $\theta_{01} $, $\theta_{02}$, $\theta_{12}$. 
In App.~\ref{app:C} we provide the explicit structure of the state $\rho_{\Co \Ho}$ and the energy preserving unitary. 
In Fig.~\ref{fig:qutrits}, we plot the regions in parameter space $\theta_{01}$ and $\theta_{02}$ in which these probabilities turn negative (all shaded areas).  
The yellow shaded area indicates the regions where Inequality \ref{th:xftbounds} is being violated.
 This clearly shows that our method allows to detect negativities also in higher dimension.
A more detailed analysis and  a straightforward generalization to arbitrary dimension is presented in App.~\ref{app:C}.

 \section{Witnessing negativity with projective energy measurements in arbitrary dimension}
\label{sec:highD}

Testing violations of Eq.~\eqref{eq:xftbound} is a way to probe the fully quantum regime in arbitrary finite-dimensional systems. However, it has a significant drawback of being  experimentally very demanding, as $\bar{\chi}$ depends on all the entries in the initial state, as well as the details of the dynamics. 
Standard fluctuation theorem experiments, on the other hand, involve only projective energy measurements. The drawback of these is the inability to probe quantum effects due to initial quantum correlations and energy coherence. 
Can our method be adapted to obtain a scheme satisfying both desiderata, that is 1. probing the fully quantum regime 2. using only the experimental capabilities required by the traditional TPM quantum fluctuation theorems? Here we answer this question in the positive.

We consider two $d$-dimensional systems and for simplicity we assume there are no matching gaps in their energy spectrum (formally, `non-degenerate Bohr spectrum'). The most general energy-preserving unitary is then a direct product of two dimensional rotations. This allow us to derive the following result (see App.\ref{app:C} for the proof)
\begin{theorem}
	\label{thm:newbound}
	Consider two arbitrary finite-dimensional systems with non-degenerate Bohr spectrum undergoing an energy-preserving unitary. If $\pwm$ is nonnegative then
	\begin{equation}
	\label{eq:newbound}
Q \in \left[Q^{\rm TPM} -2 \lambda_-, Q^{\rm TPM} + 2\lambda_+ \right],
	\end{equation}
	where 
	\begin{align}
	\lambda_- = &  \sum_{E_{i_\Co} > E_{f_\Co}} \ptpm_{i_\Co i_\Ho \rightarrow f_\Co f_\Ho} \Delta E_{i_\Co f_{\Co}}, \\ \lambda_+ = &  \sum_{E_{i_\Co} < E_{f_\Co}} \ptpm_{i_\Co i_\Ho \rightarrow f_\Co f_\Ho} \Delta E_{f_\Co i_{\Co}}.
	\end{align}
	and $\Delta E_{nm} := E_n -E_m$. 
\end{theorem}
 
Remarkably, to confirm a violation of Eq.~\eqref{eq:newbound} and witness negativity, it suffices to have the experimental capability of performing projective energy measurements. In fact, $\lambda_-$ and $\lambda_+$ and $Q^{\rm TPM}$ can all be inferred from the data already required to test standard TPM fluctuation theorems. To infer $Q$, one can prepare $N$ samples of the initial state and use half to estimate the average energy before the dynamics and half to estimate the average energy after the dynamics. The difference between the two is the heat $Q$. In other words,  if one can protect the initial state from decoherence,  negativity can be witnessed in any of the platforms used to test TPM fluctuation theorems.  Its worth noting that Inequality~\ref{thm:newbound} does not require local thermality. In other words, it can also be tested measuring energy exchange between two quantum systems in a generic state.

\begin{figure}[t!]
\includegraphics[width=1.03\linewidth]{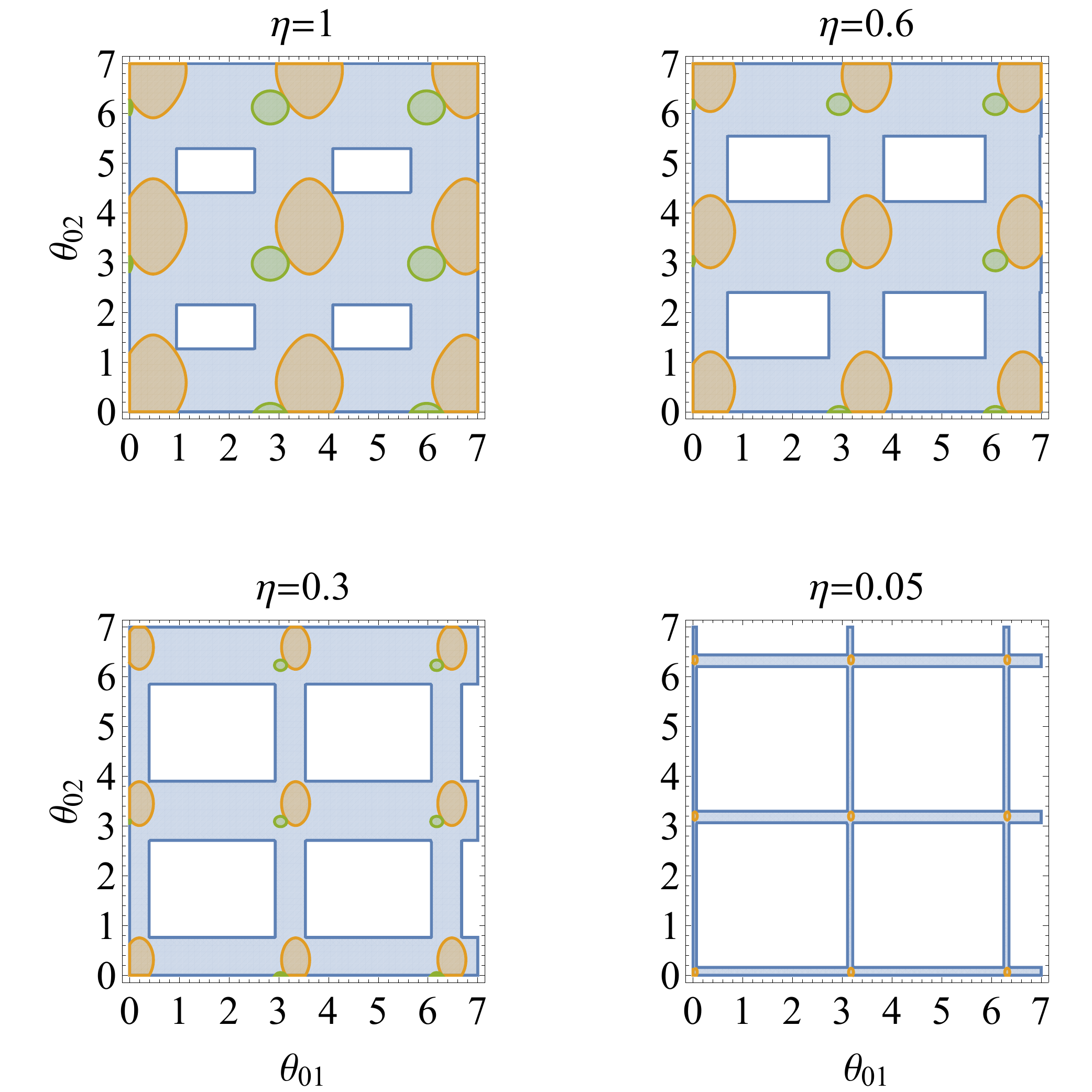}
	\caption{ 
	Negativity of the  quasiprobabilities (all shaded areas) and the violation of Inequality \ref{thm:newbound} (yellow shaded area) for different interaction protocols determined by $\theta_{01} $ and $\theta_{02} $.
The insets shows different initial states, determined by the parameter $\eta=\eta_{ij}$ for  $(i,j) = \{(1,3), (2,6), (5,7) \}$	
Parameters: $\beta_{\Ho}=0.3$, $\beta_{\Co}=1.3$, $\theta_{12}=\theta_{02}$, $E_1=1$, $E_2=1.15$,  $\xi_{ij} = 0$, $\rho_0=0.3$, $\rho_5 =0.03$, $\rho_7 =0.07$, and $\rho_8 =0.06$.}  
	\label{fig:qutrits2}
\end{figure}

To illustrate this result we consider the two qutrits case study of Sec.~\ref{subsec:case_study}.
Fig.~\ref{fig:qutrits2} presents the negativity of the quasiprobability related to heat exchange (all shaded area), and the violation of the Inequality~\ref{thm:newbound} (green and yellow areas) for different interaction protocols determined by the unitary angles $\theta_{01}$ and $\theta_{02}$.
The different insets represent different initial states, determined by the parameter $\eta=\eta_{ij}$ for  $(i,j) = \{(1,3), (2,6), (5,7) \}$ that quantifies the amount of initial quantum correlations between the states.
The yellow area represents violation of the lower bound of Inequality~\ref{thm:newbound} and corresponds to negativity in the direct flow as illustrated in the 
quantum contributions of 
Fig.~\ref{fig:heatflows} (upper right process, red arrows). The green area, on the other hand, represents violations of the upper bound of Inequality~\ref{thm:newbound}, that corresponds to negativity in the backflow as illustrated in the quantum contributions of Fig.~\ref{fig:heatflows} (lower right process, blue arrows).
The upper left inset of Fig.~(\ref{fig:qutrits2}) with $\eta=1$ corresponds to the parameters used for Fig.~\ref{fig:qutrits}. This indicates that Inequalities~\ref{th:xftbounds} and \ref{thm:newbound} can witness negativity in different regimes.  As already remarked, however, testing Inequality~\ref{thm:newbound} is much simpler.

\section{Outlook}
\label{sec:outlook}
 Recent no-go results~\cite{perarnau2017no, lostaglio2018quantum} strongly suggest that, if we are to probe the fully quantum thermodynamic regime, due to noncommutativity we need to renounce a straightforward statistical interpretation of microscopic fluctuation processes. Moving in this direction, here we introduced a natural quasiprobability for heat fluctuations whose negativity captures nonclassical effects in the strong form of contextuality. This formally implies the impossibility of describing the weak measurement scheme probing $\pwm$ as the result of any underlying classical stochastic process.  Differently from previous approaches, we showed that negativities in the heat fluctuations have a thermodynamic interpretation and that certain heat flows between two locally thermal states at different temperatures can only happen in their presence.  To prove the latter, we developed families of heat flow inequalities which can be only violated in the presence of negativities. We used these tools to witness nonclassicality by analysing data collected in a recent two-qubit experiment. We also presented heat flow inequalities for generic finite-dimensional systems whose testing only requires the control already needed to test standard TPM fluctuation theorems, with the advantage that one is testing the fully quantum regime. 
 
 Moving forward, we propose that a quasiprobability approach to fluctuations can advance our understanding of the thermodynamic behavior of realistic devices and the study of heat flows in more complex networks. Classically unachievable heat flow configurations can also be used as a witness of nonclassicality in thermal machines and could unlock performance boosts in thermodynamic protocols.  We also expect the same ideas developed here can  be applied  to work fluctuations as well. Building on the seminal paper of Allahverdyan \cite{allahverdyan2014nonequilibrium}; this may provide the most direct experimental test of the framework developed here. 
 
 More broadly, this work also opens up the possibility of applying to nonequilibrium thermodynamics the extensive technical tools based on quasiprobability representations developed to identify quantum effects and advantages in quantum optics, quantum foundations, quantum computing and condensed matter physics.

\begin{acknowledgments}
We thank David Jennings and Karen Hovhannisyan for helpful discussions. ML thanks Eran Rabani for the kind hospitality during a visit to UC Berkeley, where part of this work was completed.  ML acknowledges financial support from the the European Union's Marie Sklodowska-Curie individual Fellowships (H2020-MSCA-IF-2017, GA794842), Spanish MINECO (Severo Ochoa SEV-2015-0522 and project QIBEQI FIS2016-80773-P), Fundacio Cellex and Generalitat de Catalunya (CERCA Programme and SGR 1381). This work is part of the 'Photonics at Thermodynamic Limits' Energy Frontier Research Center funded by the U.S. Department of Energy, Office of Science, Office of Basic Energy Sciences under Award Number DE-SC0019140.
\end{acknowledgments}

\bibliography{Bibliography_thermodynamics}

\appendix

\section{Estimating the Margenau-Hill quasiprobability}
\label{app:A}
\subsubsection{Traditional scheme} 

The  scheme for measuring $\pwm\itf $ consists of a weak measurement at the start of the protocol and a projective measurement at the end. Specifically, define a family of measurement schemes where at the start the system projectors $\Pi^{i_\Co i_\Ho}$ are coupled for a unit time to the momentum $P$ of a one-dimensional pointer device through the interaction Hamiltonian $ \Pi^{i_\Co i_\Ho} \otimes P$. The pointer is initially in a pure state $ (\pi s^2)^{-1/4} \int dx e^{-x^2/{2s^2}}\ket{x}$. Then the dynamics, represented by some unitary $U$, takes place on CH. At the end a final projective energy measurement is performed on CH and outcome $(f_{\Co}, f_{\Ho})$ is observed with probability  $q_{f_{\Co} f_{\Ho}}$. 

One can then verify that the expected position of the pointer \emph{given} that some energies $(f_{\Co}, f_{\Ho})$ are observed in the final energy measurement, denoted by $\langle X \rangle_{|f_{\Co} f_{\Ho}}$, can be directly related to $\pwm\itf$ in the weak measurement limit $s \rightarrow \infty$ (large initial spread of the pointer):
\begin{equation}
\langle X \rangle_{|f_{\Co} f_{\Ho}} q_{f_{\Co} f_{\Ho}}\stackrel{s \rightarrow \infty}{\longrightarrow} \pwm\itf.
\end{equation}
The same expression gives $\ptpm_{i_\Co i_\Ho \rightarrow f_\Co f_\Ho}$ if $s \rightarrow 0$ (very sharp pointer). For a derivation see for example Appendix~B of Ref.~\cite{lostaglio2018quantum}. In this sense, the protocol probing $\pwm\itf$ can be understood as a `minimally invasive' version of the standard TPM scheme.

\subsubsection{Qubit probe}

An alternative scheme to estimate $\pwm\itf$ uses only a qubit pointer rather than a continuous system~\cite{wu2009weak, kunjwal2018anomalous} and, for completeness,  will be briefly discussed here. Take a qubit ancilla in a state $\ket{\psi_\epsilon} = \cos \epsilon \ket{0} - \sin \epsilon \ket{1}$. Couple system and ancilla through the unitary $V = \Pi^{i_\Co i_\Ho \perp} \otimes \id + \Pi^{i_\Co i_\Ho } \otimes \sigma_z$ (generated by the Hamiltonian $H_{\rm int} = \Pi^{i_\Co i_\Ho} \otimes \ketbra{1}{1}$ by $V=e^{-i g H_{\rm int} t}$, setting $t=\pi/g$). The ancilla is then measured in the $\ket{\pm} = (\ket{0} \pm \ket{1})/\sqrt{2}$ basis. Denoting the two outcomes by $\pm$, the corresponding Kraus operators on the system are
\begin{equation}
M_{\pm} = \bra{\pm} V \ket{\psi_\epsilon} = \frac{\cos \epsilon }{\sqrt{2}} \id \pm \frac{\sin \epsilon}{\sqrt{2}} (\Pi^{i_\Co i_\Ho} - \Pi^{i_\Co i_\Ho \perp}),
\end{equation}
with the correspondent positive operator valued measurement (POVM) on the system being
\begin{equation}
E_{\pm} = M^\dag_{\pm}M_{\pm} = (1 \mp \sin 2\epsilon) \frac{\id }{2} \pm \sin 2\epsilon \Pi^{i_\Co i_\Ho}. 
\end{equation} 
Denote the joint probability of observing outcome $+$ on the pointer and outcome $(f_{\Co}, f_{\Ho})$ on the system by $q_{f_{\Co} f_{\Ho},+}(\epsilon)$. Define, with obvious notation, also $q_{f_{\Co} f_{\Ho},-}(\epsilon)$. Then, if $\Delta q_{f_{\Co} f_{\Ho}}(\epsilon): = q_{f_{\Co} f_{\Ho},+}(\epsilon) - q_{f_{\Co} f_{\Ho},-}(\epsilon)$,
\begin{align*}
\Delta q_{f_{\Co} f_{\Ho}}(\epsilon) & =  \tr{}{\Pi^{f_\Co f_\Ho}(\tau) \otimes \ketbra{+}{+} V(\rho_{\Co \Ho} \otimes \ketbra{\psi_\epsilon}{\psi_\epsilon})V^\dag} \\
& - \tr{}{\Pi^{f_\Co f_\Ho}(\tau) \otimes \ketbra{-}{-} V(\rho_{\Co \Ho} \otimes \ketbra{\psi_\epsilon}{\psi_\epsilon})V^\dag} \\
& = \sin 2 \epsilon  ( 2 \pwm\itf - p_{f_\Co f_\Ho}),
\end{align*}
where  $\Pi^{f_\Co f_\Ho}(\tau) = U^\dag \Pi^{f_\Co f_\Ho} U$ and $p_{f_\Co f_\Ho} = \tr{}{\Pi^{f_\Co f_\Ho}(\tau) \rho_{\Co \Ho}}$ is the  probability of observing energy outcomes $(f_\Co f_\Ho)$ if no measurement scheme is performed ($V= \id$). Clearly, 
\begin{equation}
\pwm\itf = \frac{\Delta q_{f_{\Co} f_{\Ho}}(\epsilon)}{2 \sin 2\epsilon} + \frac{p_{f_\Co f_\Ho} }{2}.
\end{equation}
This gives a way to reconstruct $\pwm\itf $ from the joint statistics on pointer and system in the abovementioned measurement scheme, together with the probabilities  $p_{f_\Co f_\Ho}$ that can be inferred from a second experiment where no measurement scheme is applied. Note that, given the knowledge of $\epsilon$ from the initialization of the ancilla, one can reconstruct $\pwm\itf$ without taking the limit $\epsilon \rightarrow 0$.

\section{Heat flow between Two qubits}
\label{app:B}

\subsubsection{Reduction of the parameters}

The most general two-qubit unitary satisfying energy conservation takes the form
\begin{equation}
\label{eq:U}
U = \left(\begin{array}{cccc}
1 & 0 & 0 & 0\\
0 & e^{i (\kappa + \lambda) } \cos \theta  &  - e^{i (\kappa - \phi) }  \sin \theta & 0\\
0 &e^{i (\kappa + \phi) } \sin \theta   &e^{i (\kappa - \lambda)}  \cos \theta   & 0\\
0 & 0 & 0 &1
\end{array}\right).
\end{equation}
Let $\mathcal{D}$ be the global dephasing operation in the energy basis, 
\begin{equation}
\mathcal{D}(\cdot) = \sum_{E \in {\rm spec}(H_\Co + H_\Ho)} \Pi_{E} (\cdot)\Pi_E, 
\end{equation}
where spec$(X)$ is the spectrum of $X$ and $\Pi_{E}$ is the projector on the eigenspace of energy $E$. One can verify that, for all $U$ with $[U,H_\Co + H_\Ho] = 0$, both $p^{\rm TPM}$ and $p^{\rm W}$ are not affected by the application of $\mathcal{D}$ on the initial state $\rho_{\Co \Ho}$. For the two qubit case this implies that, without loss of generality, we can take the state with thermal marginals to have the form
\begin{equation}
\label{eq:rho}
\rho_{\Co \Ho} = \left(\begin{array}{cccc}
P_{00} & 0 & 0 & 0\\
0 & \frac{1}{z_\Co}-P_{00} & \eta e^{i\xi} & 0\\
0 & \eta e^{-i\xi} & \frac{1}{z_\Ho}-P_{00} & 0\\
0 & 0 & 0 &\frac{z_\Co z_\Ho -z_\Co-z_\Ho}{z_\Co z_\Ho}+P_{00}
\end{array}\right),
\end{equation}
where $z_{\Co(\Ho)}=1+e^{-\beta_{\Co(\Ho)}}$. The two qubit must be resonant to have a nontrivial dynamics, due to the fact that $U$ conserves energy (we will discuss how to weaken this assumption later). We then set $H_{\Co} = H_{\Ho} = \ketbra{1}{1}$ by renormalizing the temperature.
Since $\rho_{\Co \Ho}$ has to be a valid density operator, $P_{00}$ must satisfy
\beqar
P_{00} &\leq & \frac{1}{z_{\Ho}}, \label{eq:p00}\\ 
P_{00} &\geq & \frac{ z_{\Co}+z_{\Ho}-z_{\Co} z_{\Ho}}{z_{\Co} z_{\Ho}}\label{eq:p00_min}.
\eeqar
The MH and the TPM probabilities satisfy
\begin{align}
\pwm\zto=\ptpm\zto+\eta\cos \theta \sin\theta \cos \xi, \nonumber \\ \label{eq:prob_qubits} 
\pwm\otz=\ptpm\otz-\eta\cos\theta\sin\theta \cos \xi, \\ \nonumber
\ptpm\zto = \left(\frac{1}{z_{\Co}} - P_{00}   \right) \sin^2\theta, \\
\ptpm\otz = \left( \frac{1}{z_{\Ho}} - P_{00}\right) \sin^2\theta. \label{eq:prob_qubits2} 
\end{align}

The probabilities $\pwm$ and $\ptpm$ are independent of $\kappa$, hence we set $\kappa = 0$. Furthermore, $\ptpm$ are independent of $\lambda$, $\xi$ and $\phi$, while $\pwm$ only depends upon them by a factor $\cos (\lambda + \xi + \phi)$. A simple reparametrization then allows one to set $\lambda = \phi = 0$. One can also set $\xi = 0$ by a redefinition of $\eta$. We can always take $\eta \geq 0$ by a reparametrization of $\theta$ (if $\eta \leq 0$ map $\theta \mapsto -\theta$). In the main text, we allowed $\eta \in \mathbb{R}$ for an easier comparison to the experimental parameters of Ref.~\cite{micadei2019reversing}. The final forms are
\begin{equation}
\label{eq:U_final}
U = \left(\begin{array}{cccc}
1 & 0 & 0 & 0\\
0 & \cos(\theta) & -\sin(\theta)  & 0\\
0 & \sin(\theta) & \cos(\theta) & 0\\
0 & 0 & 0 &1
\end{array}\right),
\end{equation}
\begin{equation}
\label{eq:rho_reduced}
\rho_{\Co \Ho} = \left(\begin{array}{cccc}
P_{00} & 0 & 0 & 0\\
0 & \frac{1}{z_\Co}-P_{00} & \eta  & 0\\
0 & \eta & \frac{1}{z_\Ho}-P_{00} & 0\\
0 & 0 & 0 &\frac{z_\Co z_\Ho -z_\Co-z_\Ho}{z_\Co z_\Ho}+P_{00}
\end{array}\right),
\end{equation}
where $z_{\Co(\Ho)}=1+e^{-\beta_{\Co(\Ho)}}$ and $\eta \geq 0$.

A direct computation of the heat 
\beq
\nonumber
Q := \tr{}{\rho_{\Co \Ho} H_C} - \tr{}{U\rho_{\Co \Ho}U^\dag H_C}
\eeq returns
\begin{small}
\begin{equation}
\label{eq:Qgeneral}
Q = - \eta \cos \xi \sin 2\theta + \sin^2 \theta \left( \frac{1}{1+e^{\beta_{\Co}}} -  \frac{1}{1+e^{\beta_{\Ho}}} \right).
\end{equation}
\end{small}
Note that $Q$ does not depend on $P_{00}$ and that $\rho_{\Co \Ho} \geq 0$ implies $|\eta| \leq \sqrt{(1/z_\Co - P_{00}) (1/z_\Ho - P_{00})}$. 

Concerning $Q^{\rm TPM}$,
\beq
\nonumber
Q^{\rm TPM} := \tr{}{\mathcal{D}(\rho_{\Co \Ho}) H_C}) - \tr{}{U\mathcal{D}(\rho_{\Co \Ho})U^\dag H_C}.
\eeq
 We have
\begin{equation}
\label{eq:qtpmtwoqubit}
Q^{\rm TPM} = \sin^2 \theta \left( \frac{1}{1+e^{\beta_{\Co}}} -  \frac{1}{1+e^{\beta_{\Ho}}} \right).
\end{equation}
Hence $Q^{\rm TPM} \leq 0$, i.e. for the two-qubit case no backflow exists in the TPM scheme.
Note that, as it is well known, $Q(\eta = 0) = Q^{\rm TPM}$. That is, the average heat flows of the TPM scheme coincides with that of a state from which all initial coherence has been removed.

\subsubsection{Backflow and quantum correlations in two-qubit systems}

Next we show that, for a state that is diagonal in the energy basis (even if it includes classical correlations), no heat backflow can be observed. This can be checked by simply replacing $\eta = 0$ in Eq.~\eqref{eq:Qgeneral}, which gives  
\beq
\label{eq:qtpmqubit}
Q(\eta = 0) = Q^{\rm TPM} \leq 0.
\eeq
Hence, $\eta \neq 0$ is necessary for backflow.

\subsection{Proof of Inequality 1 }
\label{subsec:proof1}
Assume we observe a direct flow, i.e. $Q<0$. From Eqs.~\eqref{eq:prob_qubits}-\eqref{eq:prob_qubits2} this implies $\theta  \neq 0$.  Since we can then restrict to $\theta \in (0, \pi)$,  Eq.~\eqref{eq:qtpmqubit} and $\beta_1 \neq \beta_2$ imply $Q^{\rm TPM} < 0$. Hence, we can define $\alpha := Q/\qtpm$. From Eqs.~(\ref{eq:prob_qubits}) we obtain
\beq
2\pwm\otz=(\alpha+1)\ptpm\otz-(\alpha-1)\ptpm\zto .
\eeq
We have $\alpha > 0$ and, due to Eq.~\eqref{eq:p00}, $\ptpm\zto > 0$. 
Then $\pwm\otz \geq 0$ implies 
\beq
\label{eq:ratio1}
\frac{\ptpm\otz}{\ptpm\zto} \geq \frac{\alpha-1}{\alpha+1}=\frac{Q-\qtpm}{Q+\qtpm}.
\eeq  
We further note that, from Eqs.~\eqref{eq:prob_qubits2},
\beq
\label{eq:ratio2}
\frac{\ptpm\otz}{\ptpm\zto} = \frac{ z_{\Co} (1 - z_{\Ho} P_{00})}{z_{\Ho} (1- z_{\Co} P_{00})} \leq  \frac{1+e^{\beta_{\Ho}}}{1+e^{\beta_{\Co}}},
\eeq  
where equality is obtained for the minimal value of $P_{00}$ in  Eq.~\eqref{eq:p00_min}. 
From Eq.~(\ref{eq:ratio1}) and Eq.~(\ref{eq:ratio2}) we have 
\beq
\frac{1+e^{\beta_{\Ho}}}{1+e^{\beta_{\Co}}} \geq \frac{Q-\qtpm}{Q+\qtpm},
\eeq
which is the statement of Inequality 1 for $Q<0$.

In a similar manner, we can repeat the calculations for $Q>0$. In this case,  
\beq
2\pwm\zto=(\alpha+1)\ptpm\zto-(\alpha-1)\ptpm\otz .
\eeq
Again, we have $\qtpm < 0$, which implies $\alpha < 0$, and $\ptpm\otz > 0$. Hence $\pwm\zto \geq 0$ implies
\beq
\label{eq:ratio3}
\frac{\ptpm\otz}{\ptpm\zto} \geq \frac{\alpha+1}{\alpha-1}=\frac{Q+\qtpm}{Q-\qtpm}.
\eeq 
From Eq.~(\ref{eq:ratio2}) and Eq.~(\ref{eq:ratio3}),
\beq
\frac{1+e^{\beta_{\Ho}}}{1+e^{\beta_{\Co}}} \leq \frac{Q+\qtpm}{Q-\qtpm} .
\eeq
This is Inequality 1 in the case $Q>0$.

\subsection{Nonideal heat exchange: Inequality 2}
\label{app:nonidealflow}

\subsubsection{Assumptions and some consequences}

We find the tolerance of Inequality~1 to imperfections.   The relevant assumptions are:
\begin{enumerate}
\item[A1] Rather than assuming that the dynamics $U$ on CH satisfies $[U, H_{\Co \Ho}] = 0$ (no work injected), we only assume that there exists some $\ut$ close to $U$ ($|| U-\ut||\leq \varepsilon$) that satisfies $[\ut, H_{\Co \Ho}] = 0$ (small work injected).  
\item[A2] In the setting where no work is injected, the two local thermal states support heat exchanges only if they are resonant. We drop this assumption and define the local Hamiltonians $H_{\Co(\Ho)}=E_{\Co(\Ho)}\ketbra{1}{1}$, where $E_{\Ho}-E_{\Co}=\delta$ is how much they are off-resonance. We assume this gap is not too large: if the average energy is $\bar{E}=\frac{1}{2}(E_{\Ho}+E_{\Co})$ 
\begin{equation}
 \Delta \equiv \frac{|\delta|}{2\bar{E}} \leq \frac{e^{\beta_\Co E_\Co} - e^{\beta_\Ho E_\Ho}}{2+ e^{\beta_\Co E_\Co} + e^{\beta_\Ho E_\Ho}} \equiv \frac{1-R}{1+R},
\end{equation}
where we defined $R$ the ratio $R = \frac{1+e^{\beta_{\Ho}E_{\Ho}}}{1+e^{\beta_{\Co}E_{\Co}}}$. 
\item[A3] Since now some work is injected, we will only consider energy flows that are far enough from zero that they cannot be simply due to work being injected directly in the local system by the driving. Our conclusions will hence be valid only for dynamics $U$ inducing energy flows
\begin{equation}
|Q(U)| \geq 2\bar{E}\varepsilon.
\end{equation}
 \end{enumerate}

Above $||\cdot||$ denotes the Shatten infinite norm
\begin{equation}
\| A \| := \max_{\ket{v} \textrm{ s.t. }  \|v \| =1} \| A \ket{v} \|.
\end{equation}
We will also use the Shatten 1-norm or trace norm
\begin{equation}
\| A \|_1 := \tr{}{\sqrt{A^\dag A}} .
\end{equation}

Let us start with some inequalities that will be useful later and then deal with the two cases (direct flow and backflow) separately. 

Using H\"older's inequality $|\tr{}{B^\dag A} |\leq \lVert B\rVert \lVert A \rVert_{1}$ and the sub-multiplicativity of the norm  $||A||_1$
\small
\begin{align}
\nonumber
&\vert \pwm\itf(U)-\pwm\itf(\ut) \vert = 
\\ \nonumber
&\Big\vert {\rm Re} \tr{}{\ud( \Pi^{f_\Co f_\Ho})U  \Pi^{i_\Co i_\Ho} \rho_{\Co \Ho}}- {\rm Re} \tr{}{\utd( \Pi^{f_\Co f_\Ho})\ut  \Pi^{i_\Co i_\Ho} \rho_{\Co \Ho}} \Big\vert 
\\ \nonumber
&\leq  \frac{1}{2} \lVert \ud-\utd \rVert \lVert \Pi^{f_\Co f_\Ho}U  (\Pi^{i_\Co i_\Ho} \rho_{\Co \Ho}+\rho_{\Co \Ho} \Pi^{i_\Co i_\Ho} ) \rVert_{1} 
\\ \nonumber
&+ \frac{1}{2} \lVert U-\ut \rVert \lVert \Pi^{f_\Co f_\Ho}\utd  (\Pi^{i_\Co i_\Ho} \rho_{\Co \Ho}+\rho_{\Co \Ho} \Pi^{i_\Co i_\Ho} ) \rVert_{1}
\\ \nonumber
&\leq  2 \lVert U-\ut \rVert \lVert \Pi^{f_\Co f_\Ho} \rVert_{1} \lVert \utd \rVert_{1}  \lVert \Pi^{i_\Co i_\Ho}\rVert_{1} \lVert \rho_{\Co \Ho} \rVert_{1} 
\\
&\leq  
  2 \lVert U-\ut \rVert \leq 2\varepsilon
  \label{eq:prob_distance}
\end{align}
\normalsize

 In a similar manner one can bound 
\beq
\label{eq:prop_tpm_distance}
\vert \ptpm\itf(U)-\ptpm\itf(\ut) \vert \leq 2\varepsilon
\eeq

Define $\tilde{p}^{\rm W}\equiv p^{\rm W}(\ut)$, $\tilde{p}^{\rm TPM}\equiv p^{\rm TPM}(\ut)$. Using Eq.~(\ref{eq:prob_qubits}) the heat under the energy preserving unitary $\tilde{U}$ can be expressed as
\beqar
\label{eq:Qut}
Q(\ut)&=&\ptwm\otz E_{\Ho} -\ptwm\zto E_{\Co}
\\ \nonumber
&=& (\ptwm\otz-\ptwm\zto)\bar{E} +\frac{1}{2}(\ptwm\otz+\ptwm\zto)\delta
\\ \nonumber
&=& (\ptwm\otz-\ptwm\zto)\bar{E} +\frac{1}{2}(\pttpm\otz+\pttpm\zto)\delta,
\\ \nonumber
\qtpm(\ut)&=& (\pttpm\otz-\pttpm\zto)\bar{E} +\frac{1}{2}(\pttpm\otz+\pttpm\zto)\delta.
\eeqar
 From Eq. (\ref{eq:prob_distance})-(\ref{eq:Qut}), 
\beqar
\nonumber
|Q(U)-Q(\ut)|\leq 2 \bar{E} \varepsilon
\\ \label{eq:Qdiff}
|\qtpm(U)-\qtpm(\ut)|\leq 2 \bar{E} \varepsilon.
\eeqar

We will now study two cases separately. 

\subsubsection{Direct flow}

Following the same procedure as in Sec.~\ref{subsec:proof1} some algebra gives
\beq
\label{eq:pozw}
2\ptwm\otz=(\alpha+1)\pttpm\otz-(\alpha-1)\left(\pttpm\zto-\frac{\delta}{2\bar{E}}(\pttpm\otz+\pttpm\zto)\right). 
\eeq
Using assumption A3 and the fact that $Q(U) \leq 0$ (direct flow)
\begin{equation}
Q(U)+2\bar{E}\varepsilon\le 0.
\end{equation}
This and Eq.~\eqref{eq:Qdiff} imply $Q(\ut)\le 0$, and so $\alpha:= Q(\ut)/Q^{\rm TPM}(\ut) \geq 0$.
Assuming $\pwm\geq 0 $, Eq.~\eqref{eq:prob_distance} implies $2\ptwm\otz + 4\varepsilon \geq 0 $. This allows us to derive from Eq. (\ref{eq:pozw}) the inequality
\beq
\frac{\pttpm\otz}{\pttpm\zto} -\frac{(\alpha-1)}{\alpha+1}+\frac{(\alpha-1)}{\alpha+1}\frac{\delta}{2\bar{E}}\left(1+ \frac{\pttpm\otz}{\pttpm\zto}\right) +\frac{4\varepsilon}{(\alpha +1)\pttpm\zto}\geq 0.
\eeq
Recalling from Eq.~\eqref{eq:ratio2} that $ \frac{\pttpm\otz}{\pttpm\zto} \leq \frac{1+e^{\beta_{\Ho}E_{\Ho}}}{1+e^{\beta_{\Co}E_{\Co}}}\equiv R$,
\begin{widetext}

\beqar
Q(\tilde{U}) &\geq & \frac{1+R-\Delta(1+R)}{1-R-\Delta(1+R)}\qttpm +4\varepsilon\frac{1-\Delta(1+R)}{1-R-\Delta(1+R)}\frac{\qttpm}{\pttpm\zto}\\ \nonumber
&\geq & \frac{1+R-\Delta(1+R)}{1-R-\Delta(1+R)}\qttpm -4\bar{E}\varepsilon\frac{(1-\Delta(1+R))(1+\Delta)}{1-R-\Delta(1+R)}.
\eeqar

In the first inequality we used assumption A2 and in the second we used the relation $\qttpm /\pttpm\zto \geq -\bar{E}(1+\Delta)$.
Finally, expressing the results for the non ideal heat exchange, we apply Eq.~(\ref{eq:Qdiff}) to the last inequality and obtain for the direct flow   
\beq
Q(U) \geq  \frac{1+R-\Delta(1+R)}{1-R-\Delta(1+R)}Q^{\rm TPM}(U) -4\bar{E}\varepsilon\frac{(1-\Delta(1+R))(2-\Delta)}{1-R-\Delta(1+R)}.
\eeq
Note that also this weaker inequality holds:
\beq
\label{eq:direct_nonideal}
Q(U) \geq  \frac{1+R-\Delta(1+R)}{1-R-\Delta(1+R)}Q^{\rm TPM}(U) -4\bar{E}\varepsilon\frac{2+ \Delta(1+R)}{1-R-\Delta(1+R)}.
\eeq
\end{widetext}

\subsubsection{Backflow} 
Following a procedure as in Sec.~\ref{subsec:proof1}, some algebra gives
\beq
2 \ptwm\zto = (\alpha+1)\pttpm\zto -(\alpha-1)\left(\pttpm\otz + \frac{\delta}{2 \bar{E}}(\pttpm\zto+\pttpm\otz) \right),
\eeq
Using assumption A3 and the fact that $Q(U) \geq 0$ (backflow)
\begin{equation}
Q(U)-2\bar{E}\varepsilon \geq 0.
\end{equation} 
This and Eq.~\eqref{eq:Qdiff} imply that $Q(\ut) \geq 0$, and $\alpha~:=~Q(\ut)/Q^{\rm TPM}(\ut) \leq 0$. Assuming $\pwm \geq 0$, we have  $2\ptwm\zto + 4\varepsilon \geq 0 $. Ths relation gives
\beq
\frac{\pttpm\otz}{\pttpm\zto} -\frac{\alpha+1}{\alpha-1}+\frac{\delta}{2\bar{E}}\left(1+ \frac{\pttpm\otz}{\pttpm\zto}\right) +\frac{4\varepsilon}{(\alpha -1)\pttpm\zto}\geq 0.
\eeq
As before, since $ \frac{\pttpm\otz}{\pttpm\zto} \leq R$ and using A2, some algebra gives
\begin{widetext}

\beqar
\qt &\leq & - \frac{1+R+\Delta(1+R)}{1-R-\Delta(1+R)}\qttpm -4\varepsilon\frac{1}{1-R-\Delta(1+R)}\frac{\qttpm}{\pttpm\zto}\\ \nonumber
&\leq & \frac{1+R+\Delta(1+R)}{1-R-\Delta(1+R)}\qttpm +4\bar{E}\varepsilon\frac{1-R+\Delta(1+R)}{1-R-\Delta(1+R)},
\eeqar

where we used the relation $-\qttpm/\pttpm\zto\leq \bar{E}(1~-~R+\Delta(1+R))$.
Applying Eq.~(\ref{eq:Qdiff}),

\beq
\label{eq:back_nonideal}
Q(U) \leq  -\frac{1+R+\Delta(1+R)}{1-R-\Delta(1+R)}Q^{\rm TPM}(U)+4\bar{E}\varepsilon\frac{(2-R+\Delta(1+R))}{1-R-\Delta(1+R)}.
\eeq 
\end{widetext}

While the two separate inequalities are stronger, the direct and back flows inequalities for non ideal heat exchange in Eqs.~(\ref{eq:direct_nonideal}) and (\ref{eq:back_nonideal}) can also be cast into a compact, symmetric representation:
\beq
|Q| \leq \frac{1+R+\Delta(1+R)}{1-R-\Delta(1+R)}|\qtpm | +4\bar{E}\varepsilon\frac{2+\Delta(1+R)}{1-R-\Delta(1+R)}.
\eeq

\subsection{Nonideal heat exchange: example} 
\begin{figure}[t!]
	\includegraphics[width=0.85\linewidth]{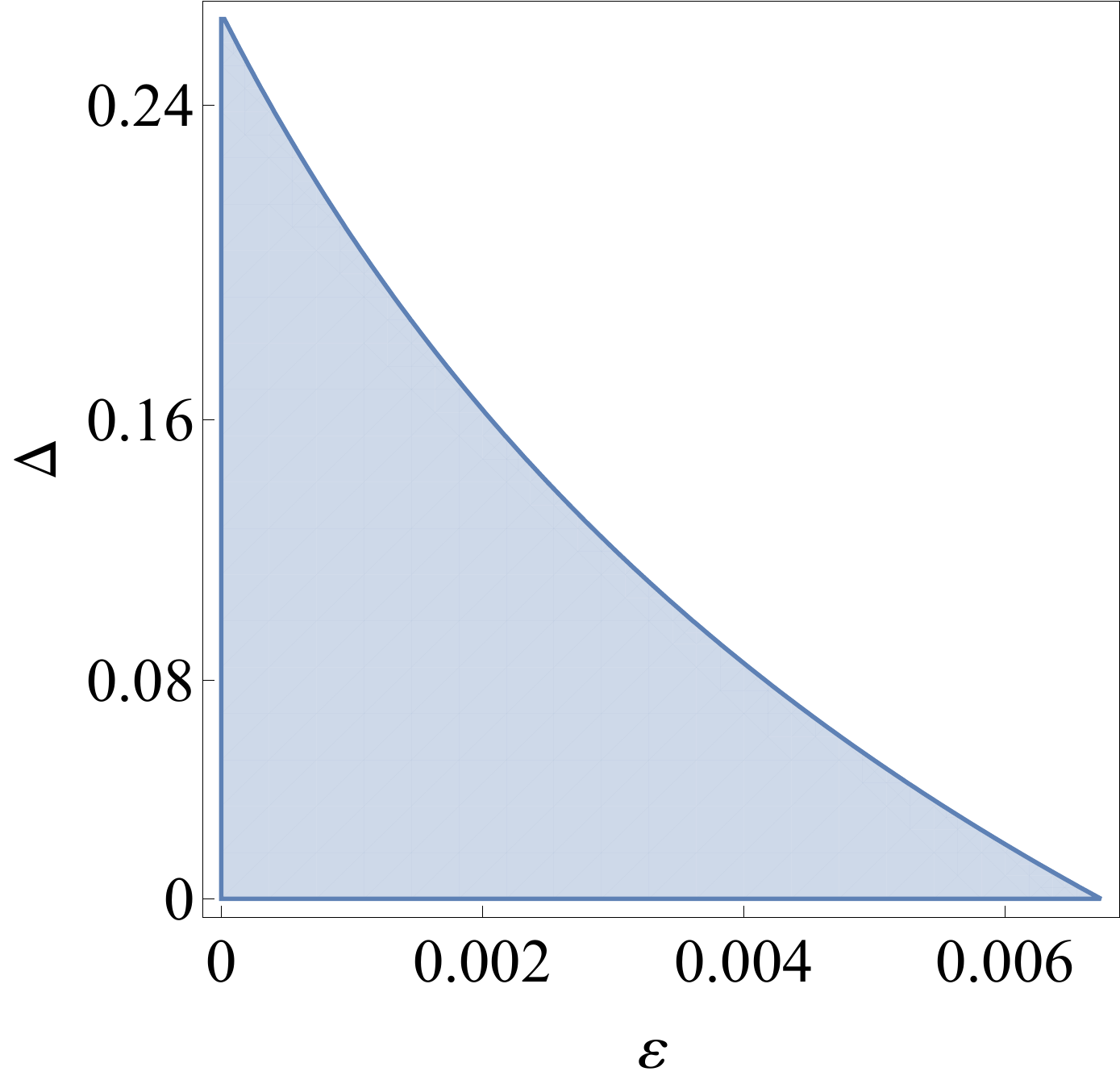}
	\caption{ The shaded area indicates the parameter regime in which Inequality~2 can be violated, i.e. negativity can be detected for non-ideal heat exchange processes. Here:  $\omega_{\Co}=1.2$kHz, $J=220$Hz, $T_{\Co}=3.48 \cdot 10^{-8}$K, $T_{\Ho}=1.74 \cdot 10^{-7}$K , $\gamma=-0.19$, and $t=4$ms. } 
	\label{fig:nonideal}
\end{figure}

 We study the violation of Inequality~2. Our starting point is the setup considered in the experiment (see Methods) with two modifications. First, we add a small perturbation, such that the unitary is no longer energy preserving. In particular, we add an interaction Hamiltonian of the form $J_{x} \sigma^{\Ho}_x \sigma^{\Co}_x$ , where $J_x$ will determine the distance between the unitaries $\varepsilon$. Second, the two qubits have different Hamiltonians that are characterized by $\Delta=(\omega_{\Ho}-\omega_{\Co})/(\omega_{\Ho}+\omega_{\Co})$. In Fig.~\ref{fig:nonideal} we present the tolerance of our inequality to non-ideal heat exchange. The shaded area indicates the violation of Inequality~2 as function of $\varepsilon$ and $\Delta$.  

\section{Extensions to arbitrary finite-dimensional systems}
\label{app:C}

\subsection{Proof of the Quantum XFT}

Here we provide a proof of the quantum XFT in the main text.
Traditionally, fluctuation theorems are derived by considering the ratio of the forward and backward process probabilities \cite{jarzynski2004classical, jennings2012exchange}. Hence, we consider
\begin{equation}
\label{eq:intermediate}
\frac{\pwm\itf}{\ptwm\fti}= \frac{{\rm Re} \tr{}{\Ud( \Pi^{f_\Co f_\Ho})  \rho^{i_\Co i_\Ho} }}
{{\rm Re} \tr{}{\U( \Pi^{i_\Co i_\Ho})  \rho^{f_\Co f_\Ho} }}
e^{-\Delta I -\Delta\beta \Delta E_{ i_{\Co} f_{\Co}} },
\end{equation}
where   
$\ptwm\fti:={\rm Re} \tr{}{\U( \Pi^{i_\Co i_\Ho})  \Pi^{f_\Co f_\Ho} \rho_{\Co \Ho}}$, $\rho^{i_\Co i_\Ho}=\frac{\Pi^{i_\Co i_\Ho}\rho_{\Co \Ho}}{\tr{}{\Pi^{i_\Co i_\Ho}\rho_{\Co \Ho}}}$ and $\Delta I := I_{f_{\Co} f_{\Ho}}-I_{i_{\Co} i_{\Ho}}$ with
\beq
I_{i_{\Co} i_{\Ho}}=\log \frac{\tr{}{\Pi^{i_\Co i_\Ho} \rho_{\Co \Ho}} }{\tr{}{\Pi^{i_\Co} \rho_{\Co}} \tr{}{\Pi^{ i_\Ho} \rho_{\Ho}} },
\eeq
the elements of the classical mutual information. Note that we assumed $E_{i_\Co} - E_{f_\Co} \approx E_{f_\Ho} - E_{i_\Ho}$ for all nonzero $\pwm\itf$.  One can then obtain the quantum XFT from Eq.~\eqref{eq:intermediate} by multiplying both sides by $\ptwm\fti e^{\Delta I +\Delta\beta \Delta E_{ i_{\Co} f_{\Co}} }$ and summing over all indexes.

\subsection{Alternative bound to Inequality~3 }
Here we derive an alternative bound (Inequality 4) for the detection of negativity. However, we note that, for all the case studies considered, we found Inequality~3 to be superior to Inequality~4.
\par
A direct calculation of  $\sum p_{\itf} e^{\Delta \beta \Delta E{i_{\Co} f_{\Co}}}$ leads to the following
\beqar
\label{eq:ave_exp}
\langle e^{\Delta \beta \Delta E{i_{\Co} f_{\Co}}} \rangle &=& 1+ J\\ \nonumber 
&=& 1 + {\rm Re} \tr{}{\mathcal{U}^\dag(\rho_{\Co} \otimes \rho_{\Ho})(c(\rho) + q(\rho ))} 
\eeqar
where
\begin{equation}
c(\rho) = \sum_{i_{\Co} i_{\Ho}} \frac{\Pi^{i_{\Co} i_{\Ho}} (\rho_{\Co \Ho} - \rho_{\Co} \otimes \rho_{\Ho}) \Pi^{i_{\Co} i_{\Ho}}}{\tr{}{\Pi^{i_{\Co}} \rho_{\Co}}\tr{}{\Pi^{i_{\Ho}} \rho_{\Ho}}},
\end{equation}
\begin{equation}
q(\rho) = \sum_{i_{\Co} i_{\Ho}} \frac{\Pi^{i_{\Co} i_{\Ho}} \rho_{\Co \Ho}  \Pi^{  {i_{\Co} i_{\Ho}} \perp}}{\tr{}{\Pi^{i_{\Co}} \rho_{\Co}}\tr{}{\Pi^{i_{\Ho}} \rho_{\Ho}}}.
\end{equation}
Note that $c(\rho_{\Co \Ho}) = 0$ if the populations of $\rho_{\Co \Ho}$ coincide with those of $\rho_{\Co} \otimes \rho_{\Ho}$, even if $\rho_{\Co \Ho}$ has coherence. This is why we use the notation ``$c$'' to indicate ``classical correlations''. On the other hand, $q(\rho_{\Co \Ho})=0$  whenever there is no coherence, even in the presence of classical correlations, hence the notation ``$q$''.
The term $1$ in Eq.~(\ref{eq:ave_exp}) was obtained noting that 
\beqar
{\rm Re} \tr{}{\mathcal{U}^\dag(\rho_{\Co} \otimes \rho_{\Ho})\left(\sum_{i_{\Co} i_{\Ho}} \frac{\Pi^{i_{\Co} i_{\Ho}} (\rho_{\Co} \otimes \rho_{\Ho}) \Pi^{i_{\Co} i_{\Ho}}}{\tr{}{\Pi^{i_{\Co}} \rho_{\Co}}\tr{}{\Pi^{i_{\Ho}} \rho_{\Ho}}}\right)}  \nonumber \\
={\rm Re} \tr{}{\mathcal{U}^\dag(\rho_{\Co} \otimes \rho_{\Ho})(\sum_{i_{\Co} i_{\Ho}} \Pi^{i_{\Co} i_{\Ho}})} =1. \nonumber
\eeqar
\begin{theorem}
 \label{th:xftbounds_alternative} 
Let $\rho_{\Co \Ho}$ be an arbitrary finite-dimensional system with thermal marginals ($\beta_{\Co} \neq \beta_{\Ho}$) and $U$ an energy-preserving unitary, $[U, H_\Co + H_\Ho] = 0$. If $\pwm$ is nonnegative, 
\begin{equation}
\label{eq:xftbound_alternative}
Q \leq (\Delta \beta)^{-1}\log (1 +J).
	\end{equation}
	\end{theorem}
The result is an application of Jensen's inequality to Eq.~\eqref{eq:ave_exp}. Using norm inequalities it can be shown that 
\begin{equation}
J \leq  \| c(\rho_{\Co \Ho})\| +  \| q(\rho_{\Co \Ho})\|,
\end{equation}
so the r.h.s is finite.
\subsection{Heat flow between two qudits}
\begin{figure}
	\includegraphics[width=0.85\linewidth]{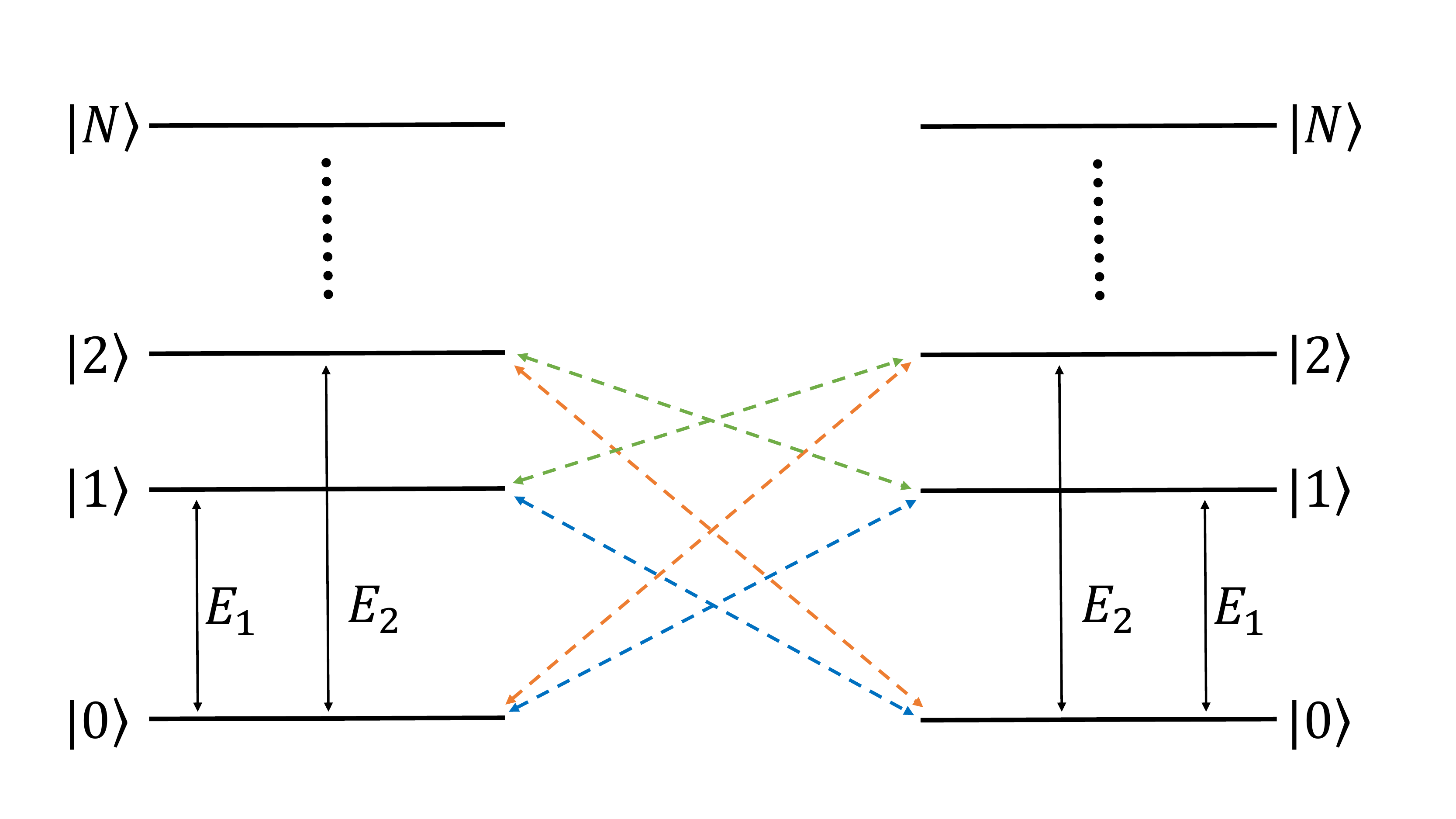}
	\caption{Schematic energy level of the two qudits and the couplings between the two.  } 
	\label{fig:qudit_schem}
\end{figure} 
To study the  heat flow between two qudits we first consider the case of two qutrits, which can then be generalized easily to $d$ dimensional systems. Setting $E_0 = 0$, we will take $H_{\Co} = H_{\Ho} = \sum_{n=1}^{2} E_n \ketbra{n}{n}$ and assume no degeneracy of the energy gaps (the `Bohr spectrum') throughout this section (see Fig.~\ref{fig:qudit_schem}).
By same reasoning as in the two qubit scenario, for the purpose of studying heat flows we can reduce a general two-qutrit state to the form,
\begin{equation}
\label{eq:rho_qutrits}
\rho_{\Co \Ho} = \left(\begin{array}{ccccccccc}
\rho_0 & 0 & 0 & 0 & 0 & 0 & 0 & 0 & 0\\
0 & \rho_1 & 0 & \rho_{13} & 0 & 0 &0 & 0 & 0\\
0 & 0 & \rho_2 & 0 & 0 & 0 &\rho_{26} & 0 & 0\\
0 & \rho_{31} & 0 & \rho_3 & 0 & 0 &0 & 0 & 0\\
0 & 0 & 0 & 0 & \rho_4 & 0 &0 & 0 & 0\\
0 & 0 & 0 & 0 & 0 & \rho_5 &0 & \rho_{57} & 0\\
0 & 0 & \rho_{62} & 0 & 0 & 0 & \rho_6 & 0 & 0\\
0 & 0 & 0 & 0 & 0 & \rho_{75} & 0 & \rho_7 & 0\\
0 & 0 & 0 & 0 & 0 & 0 & 0 & 0 & \rho_8\\
\end{array}\right),\\
\end{equation}
where $\rho_{ij}=\rho^{*}_{ji}=\eta_{ij}e^{-i\xi_{ij}}\sqrt{\rho_i \rho_j}$, with \mbox{$\eta_{ij} \in [0,1]$}, $\xi_{ij}\in \mathbb{R}$ and we use the natural labeling $(00,01,02,10,11,12,\dots) \equiv (0,1,2,3,4,5,\dots)$.  Imposing the constraints $\tr{}{\rho_{\Co\Ho}}=1$ and $\tr{\Co(\Ho)}{\rho_{\Co\Ho}}=\rho_{\Ho(\Co)}$, with
\beq
\rho_{X}= \left(\begin{array}{ccc}
1 & 0 & 0\\
0 & e^{-\beta_X E_1} &0\\
0 & 0 & e^{-\beta_X E_2}
\end{array}
\right)\tr{}{\rho_X}^{-1},
\eeq
we are left with the free parameters $\eta_{ij}$, $\xi_{ij}$ and four undetermined populations $\rho_i$ that must comply with the nonnegativity of $\rho_{\Co\Ho}$.

The general energy preserving unitary takes the form
\begin{widetext}
\begin{equation}
\label{eq:U_qutrits}
U=\left(\begin{array}{ccccccccc}
1 & 0 & 0 & 0 & 0 & 0 & 0 & 0 & 0\\
0 & e^{i(\kappa_1 +\lambda_{01})}c_{01} & 0 & -e^{i(\kappa_1 -\phi_{01})}s_{01} & 0 & 0 &0 & 0 & 0\\
0 & 0 & e^{i(\kappa_2 +\lambda_{02})}c_{02}& 0 & 0 & 0 &-e^{i(\kappa_2 -\phi_{02})}s_{02} & 0 & 0\\
0 & e^{i(\kappa_1 +\phi_{01})}s_{01} & 0 & e^{i(\kappa_1 -\lambda_{01})}c_{01} & 0 & 0 &0 & 0 & 0\\
0 & 0 & 0 & 0 & e^{i  \kappa_3 } & 0 &0 & 0 & 0\\
0 & 0 & 0 & 0 & 0 & e^{i(\kappa_4 +\lambda_{12})}c_{12} &0 & -e^{i(\kappa_4-\phi_{12})}s_{12} & 0\\
0 & 0 & e^{i(\kappa_2 +\phi_{02})}s_{02} & 0 & 0 & 0 & e^{i(\kappa_2-\lambda_{02})}c_{02} & 0 & 0\\
0 & 0 & 0 & 0 & 0 & e^{i(\kappa_4-\phi_{12}) }s_{12} & 0 & e^{i(\kappa_4 -\lambda_{12})}c_{12} & 0\\
0 & 0 & 0 & 0 & 0 & 0 & 0 & 0 & e^{i \kappa_5}\\
\end{array}\right).
\end{equation}
\end{widetext}
Here we used the short-cut notation $c_{nm}:=\cos(\theta_{nm})$ and $s_{nm}:=\sin(\theta_{nm}t)$, where $\theta_{nm}=\theta_{mn}$,  $\phi_{nm }=\phi_{mn }$ and $\lambda_{nm}=\lambda_{mn}$ are parameters characterizing the coupling between the $n$ level of the first qutrit  and the $m$ level of second qutrit.
 Within the manifolds (01,02,12), $\kappa$ are  global phases that do not change the energy transfer transition probabilities.  
One can notice the similarity between the $U$ for the qutrits with the one of the qubits. The heat flow probabilities can be split into contributions from  independent manifolds ($01$,$02$,$12$) that possess the same structure seen in Eq.~(\ref{eq:U}).   
The generalization to higher dimensions is now straightforward, as the structure of $\rho_{\Co\Ho}$ and $U$ in equations (\ref{eq:rho_qutrits}) and (\ref{eq:U_qutrits}) is preserved. 

We can express the transition probability that is related to heat transfer for arbitrary finite dimension with non degenerate gaps as
\begin{widetext}
\beqar
\label{eq:qudit_probabilities}
\forall E_n>E_m \qquad \pwm_{nm\veryshortarrow mn}&=& \rho_{n d+m}\sin^2(\theta_{nm} )\\ \nonumber
&-&\frac{1}{2} \eta_{nm}\sqrt{ \rho_{n d+m}  \rho_{m d+n}} \sin(2 \theta_{nm})\cos(\xi_{nm}+\phi_{nm}+\lambda_{nm})\\ \nonumber
\forall E_n<E_m \qquad \pwm_{nm\veryshortarrow mn}&=& \rho_{m d+n}\sin^2(\theta_{nm} )\\ \nonumber
&+&\frac{1}{2} \eta_{nm}\sqrt{ \rho_{n d+m}  \rho_{m d+n}} \sin(2 \theta_{nm} )\cos(\xi_{nm}+\phi_{nm}+\lambda_{nm}).\\ \nonumber
\eeqar 
\end{widetext}

In a similar manner, we can calculate the transition probabilities obtained form the TPM scheme,
\beqar
\label{eq:qudit_tpm_probabilities}
\forall E_n>E_m \qquad \ptpm_{nm\veryshortarrow mn}&=& \rho_{n d+m}\sin^2(\theta_{nm} )\\ \nonumber
\forall E_n<E_m \qquad \ptpm_{nm\veryshortarrow mn}&=& \rho_{m d+n}\sin^2(\theta_{nm} ).\\ \nonumber
\eeqar 

\begin{figure}
	\includegraphics[width=0.85\linewidth]{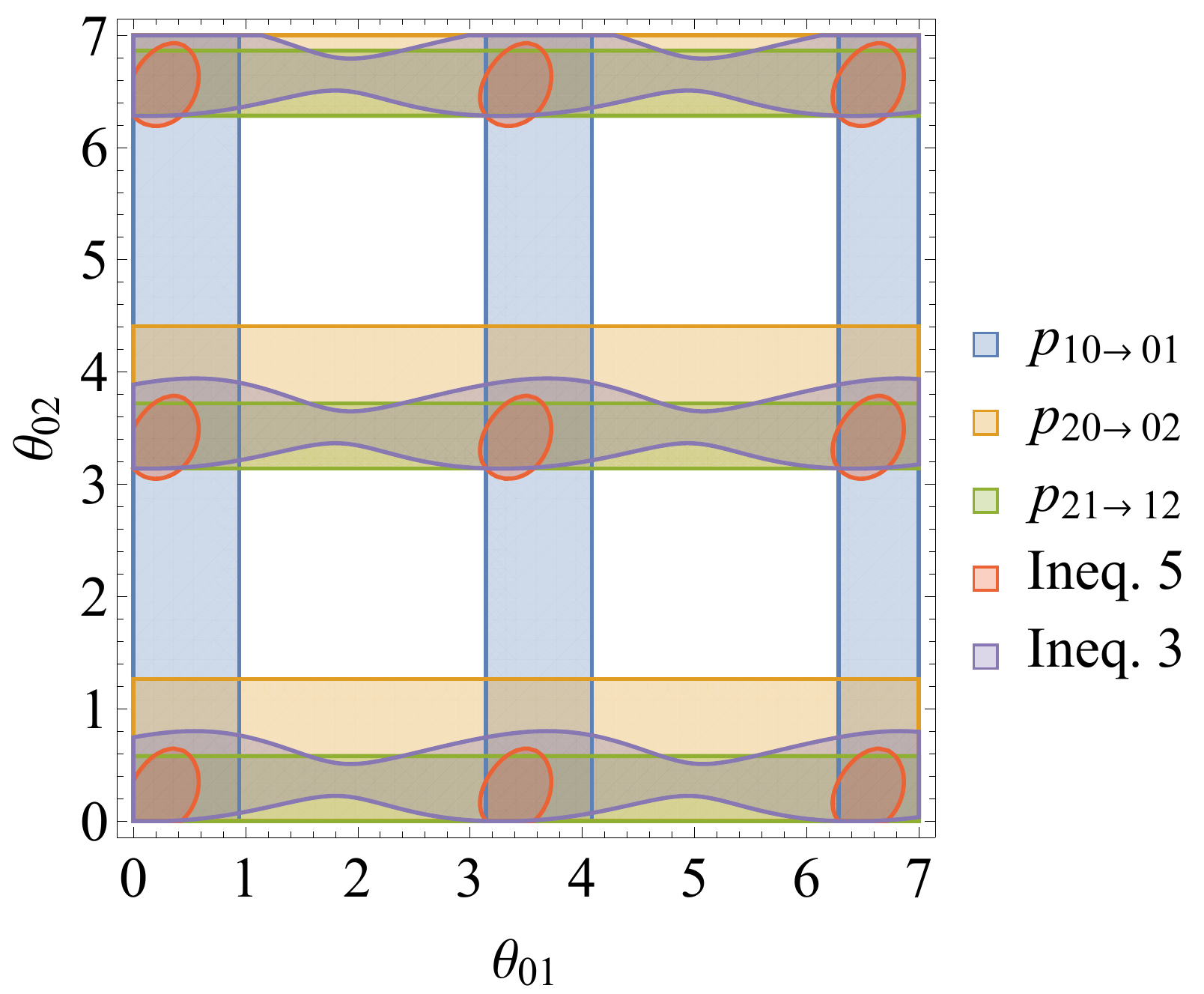}
	\caption{Direct flow negativity of the  probabilities specified in the figure and the violation of Inequality 3 and 5 for different interaction protocols determined by $\theta_{01} $ and $\theta_{02} $. Parameters: $\beta_{\Ho}=0.3$, $\beta_{\Co}=1.3$, $\theta_{12}=\theta_{02}$, $E_1=1$, $E_2=1.15$, $\xi=\phi=\lambda=0$, $\eta_{ij}= 1$ for $(i,j) = \{(1,3), (2,6), (5,7)  \}$, $\rho_0=0.3$, $\rho_5 =0.03$, $\rho_7 =0.07$, $\rho_8 =0.06$.}
	\label{fig:app:qutrits}
\end{figure} 
Here we set $E_0=0$ and ordered the levels such that $E_0 \leq E_1 \leq E_2 \leq \cdots$ and $d$ the dimension of the qudit. Next, we show that \textit{if all quasiprobabilities that contribute to direct or backflow  are negative, then necessarily $|Q|>|Q^{\rm TPM}|$. } The difference $\Delta Q=Q-Q^{\rm TPM}$ reads 
\begin{widetext}
\beq
\label{eq:deltaQ}
\Delta Q=-\sum_{E_n >E_m}\eta_{nm}\sqrt{ \rho_{n d+m}  \rho_{m d+n}} \sin(2 \theta_{nm} )  \cos(\xi_{nm}+\phi_{nm}+\lambda_{nm})(E_n-E_m). 
\eeq
\end{widetext}
For the direct flow ($Q<0$), if $\forall E_n>E_m$ we have $\pwm_{nm\veryshortarrow mn}<0$, then from Eq.~(\ref{eq:qudit_probabilities}) we immediately conclude that $\Delta Q<0$. 
For the backflow ($Q>0$), if $\forall E_n<E_m$ we have $\pwm_{nm\veryshortarrow mn}<0$, then from Eq.~(\ref{eq:qudit_probabilities}) we now have that $\Delta Q>0$, which is exactly what we wanted to prove. 
Note that Eq.~(\ref{eq:deltaQ}) also suggests a way of maximizing the difference $\Delta Q$ for a given initial state. Choosing a protocol such that $\xi_{nm}+\lambda_{nm}=-\phi_{nm}$ and $\theta_{nm}=\pm \pi/4 $ for all $n,m$, we obtain 

\beq
|\Delta Q|_{\rm max} = \sum_{E_n >E_m}\eta_{nm}\sqrt{ \rho_{n d+m}\rho_{m d+n}}(E_n-E_m).
\eeq

 In Fig.~\ref{fig:app:qutrits} we consider the two qutrits setup of Eq.~(\ref{eq:rho_qutrits}). We plot the negativity in the direct flow and compare it to the violations of Inequalities 3 and 5 for different protocols. The protocols are determined by varying $\theta_{01}$ and $\theta_{02}$.    

\subsection{Proof of inequality~\ref{thm:newbound}}

We define $\psi_{nm}=\xi_{nm}+\phi_{nm}+\lambda_{nm}$ and $\Delta E_{nm}=E_n-E_m$.
Using Eqs.~(\ref{eq:qudit_probabilities}) and (\ref{eq:qudit_tpm_probabilities}), the symmetry $\theta_{nm}=\theta_{mn}$, $\psi_{nm}=\psi_{mn}$,
and noting that 
\beqar
Q&=&\sum_{E_n>E_m}\pwm_{nm\veryshortarrow mn}\Delta E_{nm}+\sum_{E_n<E_m}\pwm_{nm\veryshortarrow mn}\Delta E_{nm} \nonumber
\\ \nonumber
&=&\sum_{E_n>E_m}\pwm_{nm\veryshortarrow mn}\Delta E_{nm}-\sum_{E_n>E_m}\pwm_{mn\veryshortarrow nm}(E_n-E_m),
\eeqar
using $\qtpm = Q(\eta=0)$ the heat exchanged can be expressed as
\small
\begin{align}
	Q = & \; \; \; \qtpm  \\ 
	- & \sum_{E_n>E_m} \eta_{nm}\sqrt{ \rho_{n d+m}  \rho_{m d+n}} \sin(2 \theta_{nm})\cos(\psi_{nm})\Delta E_{nm}. 
\end{align}
\normalsize
Using Eq.~(\ref{eq:qudit_probabilities}) and Eq.~\eqref{eq:qudit_tpm_probabilities} for $E_n>E_m$, nonnegativity implies
\beq
2\ptpm_{nm\veryshortarrow mn}>\eta_{nm}\sqrt{ \rho_{n d+m}  \rho_{m d+n}} \sin(2 \theta_{nm})\cos(\psi_{nm}).
\eeq
We conclude that 
\beq
Q>\qtpm - 2\sum_{E_n>E_m}\ptpm_{nm\veryshortarrow mn}\Delta E_{nm}.
\eeq
Violation of this inequality witness negativity in the direct flow as illustrated by upper right process of Fig.~\ref{fig:heatflows} proves the lower bound of inequality~\ref{thm:newbound}.
The upper bound on the heat can be obtained in a similar manner. To see this, note that the heat can be written as
\begin{widetext}
\beq
Q=\qtpm+ \sum_{E_n<E_m} \eta_{nm}\sqrt{ \rho_{n d+m}  \rho_{m d+n}} \sin(2 \theta_{nm})\cos(\psi_{nm})\Delta E_{mn}. 
\eeq
\end{widetext}
Using nonnegativity of Eq.(\ref{eq:qudit_probabilities}) for $E_n<E_m$ we have 
\beq
Q<\qtpm + 2\sum_{E_n<E_m}\ptpm_{nm\veryshortarrow mn}\Delta E_{mn}.
\eeq
Violation of this inequality witnesses negativity in the backflow as illustrated by lower right process of Fig.~\ref{fig:heatflows}, and proves the upper bound of inequality~\ref{thm:newbound}.

\end{document}